\documentclass[usenatbib]{mn2e}
\usepackage{graphicx}
\usepackage{ifthen}
\usepackage{amsmath}
\usepackage{url}
\usepackage{rotating}
\usepackage{xcolor}
\usepackage{caption}

\def\ltsima{$\; \buildrel < \over \sim \;$}
\def\lta{\lower.5ex\hbox{\ltsima}}
\def\gtsima{$\; \buildrel > \over \sim \;$}
\def\simgt{\lower.5ex\hbox{\gtsima}}
\def\kms{{\rm\,km \; s^{-1}}}

\def\AA{$\; \buildrel \circ \over {\rm A}$}

\def\s{\ifmmode \widetilde \else \~\fi}
\def\={\overline}

\def\spose#1{\hbox to 0pt{#1\hss}}

\def\lta{\mathrel{\spose{\lower 3pt\hbox{$\mathchar"218$}}
     \raise 2.0pt\hbox{$\mathchar"13C$}}}
\def\gta{\mathrel{\spose{\lower 3pt\hbox{$\mathchar"218$}}
     \raise 2.0pt\hbox{$\mathchar"13E$}}}
\def\Dt{\spose{\raise 1.5ex\hbox{\hskip3pt$\mathchar"201$}}}    
\def\dt{\spose{\raise 1.0ex\hbox{\hskip2pt$\mathchar"201$}}}    

\def\dotsfill{\leaders\hbox to 1em{\hss.\hss}\hfill}

\loadboldmathitalic 
\title[The Pristine Dwarf-Galaxy survey V]{The Pristine Dwarf-Galaxy survey - V. The edges of the dwarf galaxy Hercules}
\author[N. Longeard et al.] {Nicolas Longeard$^{1}$, Pascale Jablonka$^{1,2}$, Giuseppina Battaglia$^{3,4}$, Khyati Malhan$^{5}$,
\newauthor Nicolas Martin$^{5,6}$, Rub\'en S\'anchez-Janssen$^{7}$, Federico Sestito$^{8}$, Else Starkenburg$^{9}$,
\newauthor Kim A. Venn$^{8}$\\
$^{1}$ Laboratoire d'astrophysique, \'Ecole Polytechnique F\'ed\'erale de Lausanne (EPFL), Observatoire, 1290 Versoix, Switzerland\\
$^{2}$ GEPI, Observatoire de Paris, Universit\'e PSL, CNRS, Place Jules Janssen, F-92195 Meudon, France\\
$^{3}$ Instituto de Astrofísica de Canarias, San Cristóbal de La Laguna, Spain \\
$^{4}$ Department of Astrophysics, University of La Laguna, San Cristóbal de La Laguna, Spain\\
$^{5}$ Max-Planck-Institut f\"ur Astronomie, K\"onigstuhl 17, D-69117, Heidelberg, Germany\\
$^{6}$ Universit\'e de Strasbourg, CNRS, Observatoire astronomique de Strasbourg, UMR 7550, F-67000 Strasbourg, France\\
$^{7}$ STFC UK Astronomy Technology Centre, Royal Observatory, Blackford Hill, Edinburgh, EH9 3HJ, UK \\
$^{8}$ Department of Physics and Astronomy, University of Victoria, PO Box 3055, STN CSC, Victoria, BC V8W 3P6, Canada \\
$^{9}$ Kapteyn Astronomical Institute, University of Groningen, Landleven 12, 9747 AD Groningen, The Netherlands\\
}

\date{\today}
\begin{document} 
\maketitle 
\begin{abstract}
We present a new spectroscopic study of the dwarf galaxy Hercules (d $\sim 132$ kpc) with data from the Anglo-Australian Telescope and its AAOmega spectrograph together with the Two Degree Field multi-object system to solve the conundrum that whether Hercules is tidally disrupting. We combine broadband photometry, proper motions from Gaia, and our Pristine narrow-band and metallicity-sensitive photometry to efficiently weed out the Milky Way contamination. Such cleaning is particularly critical in this kinematic regime, as both the transverse and heliocentric velocities of Milky Way populations overlap with Hercules. Thanks to this method, three new member stars are identified, including one at almost $10$r$_h$ of the satellite. All three have velocities and metallicities consistent with that of the main body. Combining this new dataset with the entire literature cleaned out from contamination shows that Hercules does not exhibit a velocity gradient (d$\langle v \rangle$/d$\chi$ $ = 0.1^{+0.4}_{-0.2}$ km s$^{-1}$ arcmin$^{-1}$) and, as such, does not show evidence to undergo tidal disruption.
\end{abstract}
 

\section{Introduction}

Companion galaxies orbiting the Milky Way (MW) have been discovered at an incredible rate over the last few years, under the impulsion of various photometric surveys that are ideal to detect faint surface brightness systems (Sloan Digital Sky Survey, \citealt[SDSS]{york00}; the Panoramic Survey Telescope And Rapid Response System, \citealt[PS1]{chambers16}; the Dark Energy Survey, \citealt[DES]{abbott05}).
The faintest of them are commonly referred to as Ultra-Faint Dwarf galaxies (UFDs). 

Intensive spectroscopic observations of these very faint systems followed, mainly focusing on their dynamical and metallicity properties (e.g. , \citealt{simon_geha07}, \citealt{martin07}, \citealt{koposov11}, \citealt{walker16}, \citealt{kirby17}, \citealt{fritz19}, \citealt{chiti22}). Associated with photometric properties, these chemo-dynamical observations are instrumental to our understanding of both the nature of dark matter and the physical processes governing the evolution of baryons. These extensive observations uncovered discrepancies with respect to hydro-dynamical simulations that are yet to be solved (e.g. diversity of rotation curves, the metallicity-luminosity relation or the plane of satellites) and will only be so through a careful analysis of the properties of the faintest satellite galaxies of the MW. 

However, as more and more UFD member stars are studied, a new light has been recently shed on the faintest satellites, that focus on their potential stellar halos. If this component is highly hypothetical at such low mass, especially since their existence may relate to early mergers (\citealt{chiti21}, \citealt{tarumi21}) that are less common as one goes down the mass scale of galaxies \citep{deason22}. Of course, the observationability of these halos, should they exist in a given system that is already low surface brightness, makes their detection extremely difficult, even if recent studies have started to put them in evidence in several galaxies (\citealt{johnson20}, \citealt{pace20}, \citealt{chiti21}, \citealt{longeard22}, \citealt{qi22}). Jensen et al. (prep.) also report a few dwarf galaxies of the MW, among the 60 that went under scrutiny, for which the existence of an extended stellar halo is credible.

These efforts reveal additional layers of complexity in the kinematics and metallicity properties of the faintest galaxies. Two recent examples illustrate this complexity. The first one concerns the faint Tucana~II (Tuc~II, d $\sim 58$ kpc) satellite galaxy \citep{chiti21}. Their 7 member stars located at galactocentric distances between 2 and 9 times the Tuc~II's half light radius (r$_h$) tend to be more metal-poor than those in the galaxy central region. Should this metallicity trend be found in other UFDs, it would mean that our current view of the metallicity distribution functions (MDF) of these systems are biased, and might be lower. This would have strong implications on the galaxy formation simulations that are fine-tuned to reproduce the observed metallicities (see \citealt{sanati23} for further discussion). The second striking example of rising complexity in UFDs is the case of Boötes~I (Boo~I, d $\sim 66$ kpc). \citet{longeard22} identified 17 members in the outskirts of the satellite, including one at $\sim 4.1$ r$_h$. They measured both negative metallicity  and velocity gradients in the system. These results show that our current view of the mass functions of UFDs can also be significantly biased towards higher values, since the introduction of a velocity gradient in Boo~I dynamical modeling deflates its dynamical mass by $\sim 40$\% with respect to a simpler model with a constant systemic velocity, that is, if it the assumptions underlying its computation still hold (\citealt{wolf10}). These two recent examples perfectly illustrate the need for more spectroscopic observations in the outskirts of UFDs. If the hope for studying dwarf galaxies's halo as a whole component is extremely thin due to their predicted low surface brightness from simulations \citep{deason22}, it is still possible to detect a few large galactocentric distance stars (\citealt{yang22} for Fornax, \citealt{sestito23} for Ursa Minor, \citealt{waller23} for Coma Berenices, Ursa Major I and Boötes I). In particular, the work of \citet{waller23} focused on high-resolution spectroscopy, and abundance derivation of the outskirts of these three galaxies suggest that at least some of the stellar population of these halos can form in the inner region and migrate during the dwarf's history, while confirming that minor mergers are viable pathways to form dwarf galaxy's halos in the case of Boötes~I.

In this work, we propose to follow-up on that effort to study Hercules, a dwarf galaxy that has been the subject of speculation regarding its potential tidal disruption status. Its main properties are summarized in Table 1. This question can only be answered through the search for extra-tidal stars at large galactocentric distances that offer the largest velocity contrast with the main body. The large ellipticity and tentative velocity gradient of Hercules (\citealt{aden09}, \citealt{martin_jin10}) have warranted speculation. So far the vast majority of Hercules' known  members are located in the galaxy central region. Part of this spatial limitation is due to the fact that their identification is easier with such an observational strategy, but also because its systemic heliocentric velocity and proper motion (PM) are blended into the MW's. Therefore, the identification of new members is extremely challenging. Finding those with high confidence at large distances from the kinematic information alone is almost an impossible task.

\begin{table}
\centering
\begin{tabular}{|llll} 

 \hline
Property & Inference & Reference  \\ 
  \hline
$d_\mathrm{GC}$ (kpc) & $132.0 \pm 6.0 $ & (1) &\\ 
$r_h$ (') & $5.83 \pm 0.65$ & (1) &  \\
$r_h$ (pc) & $216 \pm 20$ & (1) &  \\
$\langle v \rangle$ (km s$^{-1}$) & $45.0 \pm 1.1$ &  (2), (3) ,(4), (5)  & \\
$\mathrm{[Fe/H]}$ & $-2.39 \pm 0.04$ &   (2), (3) ,(4), (5)   & \\

 \hline
\end{tabular}
\caption{Summary of Hercules' property. The references number correspond to the following list: (1) \citet{munoz18}, (2) \citet{simon_geha07}), (3) \citet{aden09b}, (4) \citet{deason12}, (5) \citet{gregory20}}
\label{table:1}
\end{table}

Aside from the identification of new member stars, one element has been the centre of discussion: does Hercules possess a velocity gradient that would be the telltale sign of an undergoing tidal disruption? \citet{aden09b} were the first one to detect such a gradient with their significant spectroscopic sample of $28$ red giant branch (RGB) member stars, with $d \langle \mathrm{v} \rangle /d\chi = $ $16 \pm 3$ km s$^{-1}$ kpc$^{-1}$. They immediately associated this detection with tidal disturbances in the outskirts of the UFD. \citet{martin_jin10}'s results were in line with this initial study. However, \citet{deason12} did not detect any gradient, though they make it clear that their low velocity precision, due to the poor resolution of the spectrograph used (R $\sim 2000$), may be the reason behind this non-detection. The Sections 3.1.2 and 3.1.3 of \citet{gregory20} offer detailed analyses on Hercules' velocity gradient under different assumptions. First, using only their own new $9$ members and without any prior, they did not detect any dependence of the velocity with distance. However, assuming that Hercules' velocity gradient runs along its major axis at a position angle of $-78$ deg, as detailed by \citet{martin_jin10},  they did find a gradient of $9.4^{+6.0}_{-6.3}$ km s$^{-1}$ kpc$^{-1}$. Combining these results with the spectroscopic sample of \citet{simon_geha07} yielded a similar result. Finally, \citet{kupper17} and \citet{fu19} pointed out that their predicted velocity gradients for Hercules, based on N-body simulations of the UFD, are inconsistent with the one found by \citet{aden09b}, and should be much lower, of $4.9$ km s$^{-1}$ kpc$^{-1}$ and $0.6$ km s$^{-1}$ kpc$^{-1}$ respectively. 

Furthermore, \citet{garling18} recently identified three new RR lyrae stars in Hercules outside the estimated tidal radius of Hercules, adding to the previously 9 RR Lyrae identified prior to their study \citep{musella12}. They interpreted their results as proof that some Hercules stellar material have been stripped from the system. Interestingly enough, one of their three new finding is not aligned with the major axis of the system, but with its minor axis.

More recently, \citet{errani22} shows that Hercules's velocity dispersion and size should have been affected by tides, according to their set of N-body simulations, actually lying very close to their tidal track limit (i.e. the maximum size reachable by a dwarf galaxy given its circular velocity).

The following list: \citet{aden09b}, \citet{martin_jin10}, \citet{simon_geha07}, \citet{deason12}, \citet{fu19} and \citet{gregory20} are the papers that will be referred to as ``the literature'' in the rest of this work.

The case of Hercules therefore remains open, as even the existence of a velocity gradient is still not clear, nor is its expected value should it exist. We therefore try to solve this conundrum with a re-analysis of the kinematic and metallicity properties of Hercules, using new spectroscopic observations at large galactocentric distances combined with the entire literature that include spectroscopic data.

\section{Spectroscopic observations}

\begin{figure*}
\begin{center}
\centerline{\includegraphics[width=\hsize]{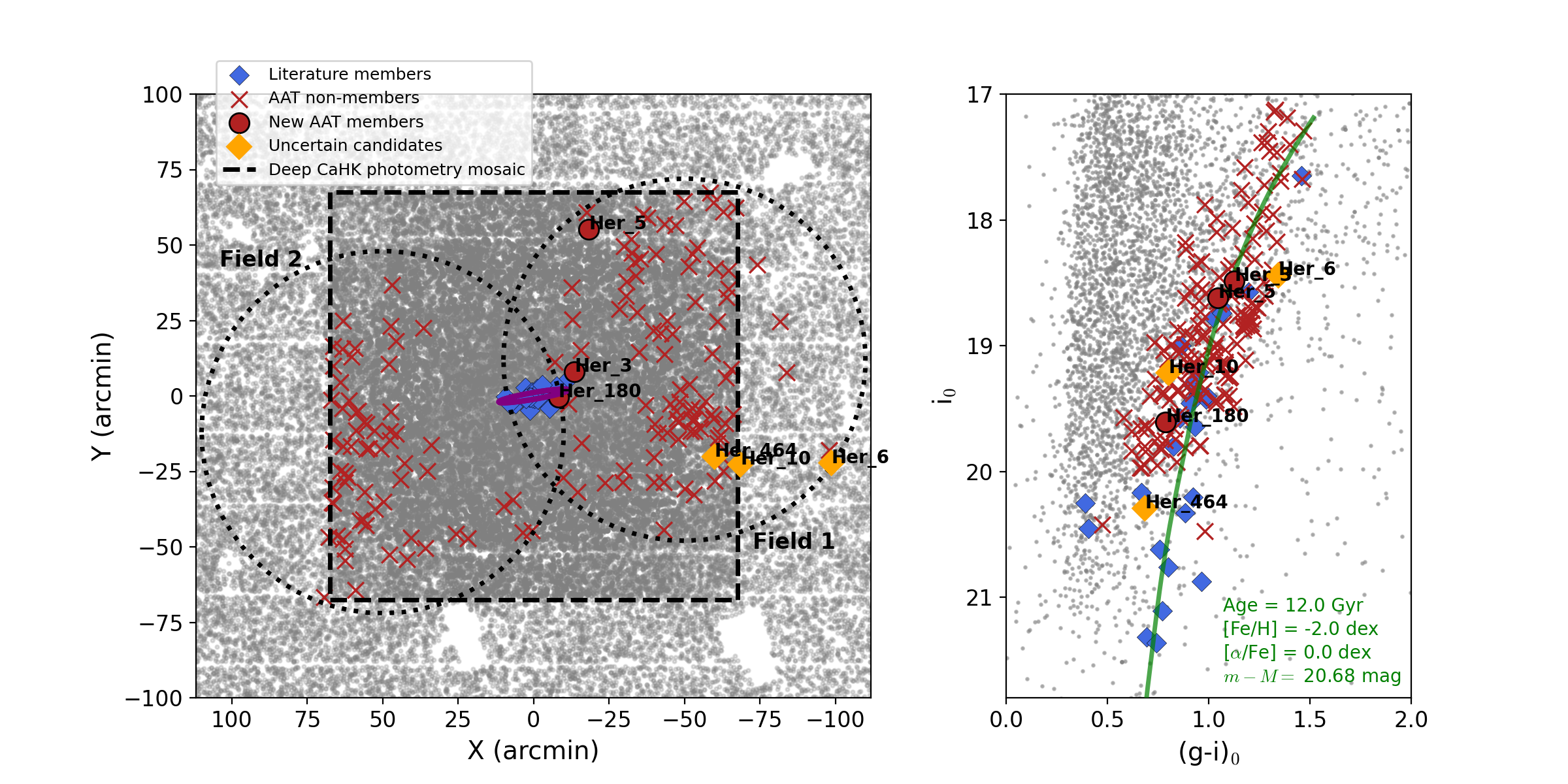}}
\caption{\textit{Left panel:} Spatial distribution of the AAT spectroscopic sample. Newly discovered members are shown as red circles, while uncertain candidates are shown as orange diamonds. Non-members from the AAT sample are shown as red crosses. Previously known members from the literature are represented as smaller blue diamonds.  The two half-light radii of Hercules as inferred by \citet[M18]{munoz18} are shown as a purple ellipse. \textit{Right panel:} CMD of our spectroscopic sample superimposed with a metal-poor Darmouth isochrone at the distance of Hercules.}
\label{field} 
\end{center}
\end{figure*}

\begin{figure}
\begin{center}
\centerline{\includegraphics[width=\hsize]{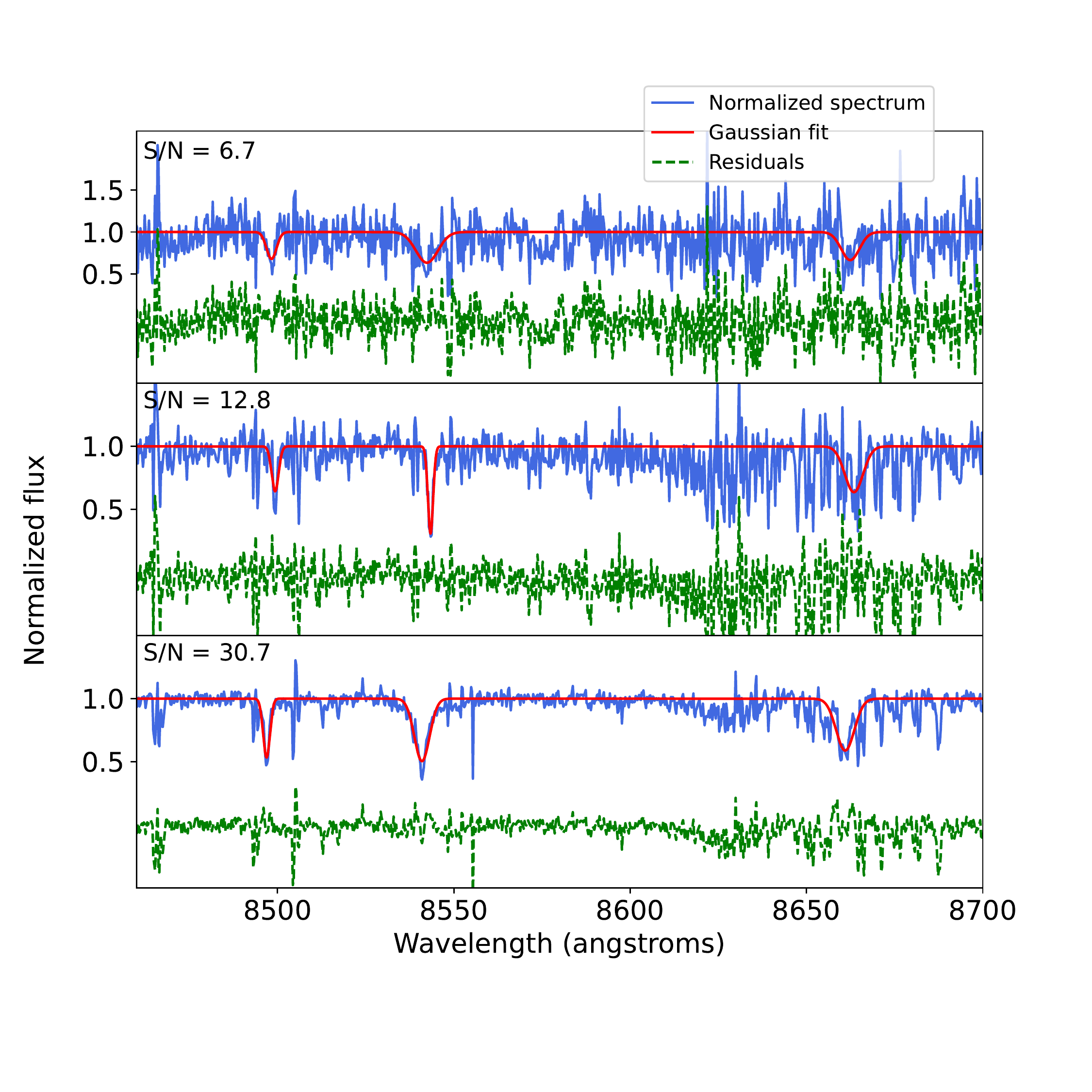}}
\caption{Example spectra of three stars in our AAT dataset centered on the calcium triplet lines. Due to the low number of new members identified, only one spectrum displayed here is a Hercules members, i.e. the second one. This spectrum is however representative of the quality of the Her' members. Each star represents respectively the low, mid and high S/N regimes. The normalised spectra are shown with solid blue lines while the fits derived from our pipeline for Gaussian profiles are shown with solid red lines. Residuals in the Gaussian cases are shown for each case below the spectra as green dashed lines. While the two first lines are properly fitted, the large sky residuals are often too large to fit the third CaT line, even for the high SNR regime. These stars have a heliocentric velocity of $31.9 \pm 3.2$, $45.5 \pm 1.4$ and $-39.9 \pm 1.1$ km s$^{-1}$ from top to bottom.}
\label{spectra_fit} 
\end{center}
\end{figure}

\begin{figure}
\begin{center}
\centerline{\includegraphics[width=\hsize]{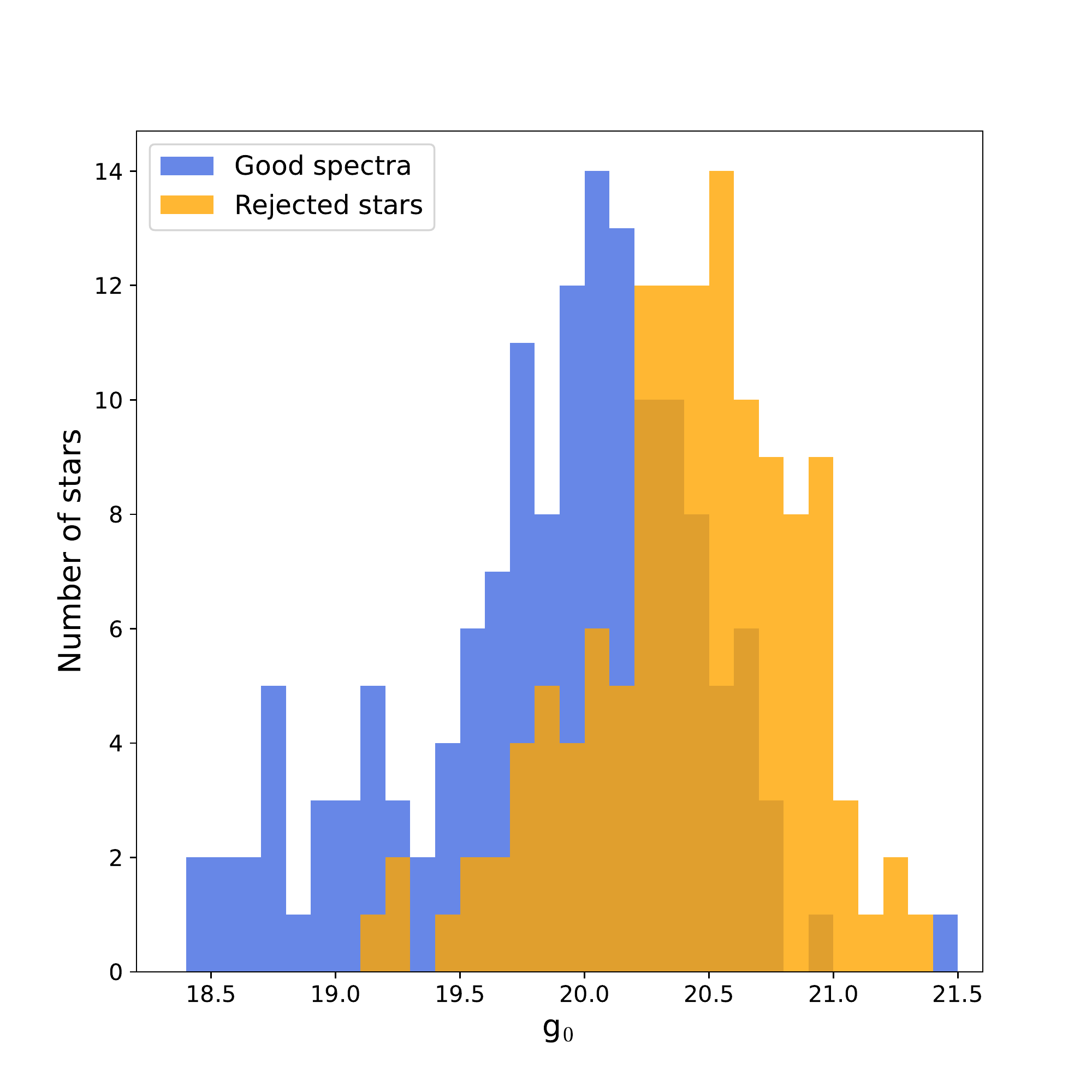}}
\caption{Histogram of the dust-corrected g magnitude for all stars with a SNR $>=$ 3 with good (blue) and poor (orange) quality spectra. Our final sample is composed of the $175$ stars in the blue sample, while the rest is rejected.}
\label{histo_rejected_mag} 
\end{center}
\end{figure}

This Section provide details on the target selection, observations and data reduction. It also introduces our pipeline to derive heliocentric velocities and equivalent widths from the spectra.

\subsection{Data selection and acquisition}

An overview of all our new targets is shown in Figure \ref{field}. This spectroscopic sample was obtained on the Anglo-Australian Telescope \citep[AAT]{lewis02} and its Two Degree Field (2dF) multi-object system \citep{cannon97} through the OPTICON program. The gratings used were 580V for low-resolution spectra in the optical (R $\sim 1300$, $3700$-$5500$\AA), and 1700D for calcium triplet spectra with a spectral resolution R of $\sim 11000$. Only the red part of the spectra (from 8400 to 8800 \AA) is used for the rest of this work. The observations were carried out on 02 and 03 May 2022. One more night was scheduled but lost due to bad weather. The 2dF spectrograph possesses $\sim 360$ science and $\sim 40$ sky and guiding fibers. During the first night, both fields benefited from 4 sub-exposures of 2400 seconds each. However, on the second night, only 2 out of the 4 sub-exposures were observed for Field 2. As a consequence, Field 2 spectra for the second night are non exploitable, and only the ones from the first night are considered. For Field 1, the final spectra are obtained by coadding the first and second nights. The total exposure time is 19200s for Field 1 and 9600s for Field 2. Spectra were gathered for 295 stars.

The two fields were placed at each extremity of the UFD along its major axis in order to find potential tidal tails. These two fields are shown in Figure \ref{field} and extend as far as $\sim 13$ half-light radii (r$_h$) of Hercules. All targets were selected based on the Pristine survey data \citep{starkenburg17}. Pristine is a photometric survey relying on a narrow-band, metallicity-sensitive photometry centered on the Calcium H\&K doublet lines taken on the Canadian France Hawaii Telescope \citep[CFHT]{boulade03}. It is successful at finding metal-poor stars against the more metal-rich MW contamination (\citealt{youakim17}, \citealt{aguado19}, \citealt{arentsen20}) and is therefore particularly suited for the UFDs metal-poor population (\citealt{longeard20}, \citealt{longeard21}, \citealt{longeard22}). For Hercules, this photometry is based on two components:

\begin{itemize}
\item A mosaic of deep Pristine images centered on Hercules, shown in Figure \ref{field}, yielding reliable photometric metallicities down to g$_0^{\mathrm{SDSS}} \sim 22.5$.
\item Shallower photometry corresponding to the Pristine main survey covering the far outskirts of the satellite, yielding reliable photometric metallicities down to g$_0^{\mathrm{SDSS}} \sim 21.5$
\end{itemize}

As illustrated in Figure \ref{field}, most of our targets were selected from the deep photometry region. Three main criteria were applied to select them:

\begin{itemize}
\item Stars located further than 0.3 mag from the best-matching Hercules isochrone (A = 12 Gyr, [Fe/H] = -2.0, [$\alpha$/Fe] = 0.0, $m - M$ = 20.68) from the Darmouth library \citep{dotter08} were discarded.
\item The photometric metallicity of all targets should be lower than $-0.5$.
\item The proper motion membership probability of all targets must be of at least $1$\%, based on the Gaia Data Release 3 \citep{gaia_dr3}. These membership probabilities are computed assuming two multivariate gaussian populations in proper motion space, for Hercules and the MW respectively. Our final systemic proper motion ($\langle \mu_\alpha^* \rangle = -0.037 \pm 0.029$ mas yr$^{-1}$, $\langle \mu_\delta \rangle = -0.365 \pm 0.043$ mas yr$^{-1}$) for the UFD is perfectly compatible with the ones of \citet{battaglia22} and \citet{mcconnachie_venn20}.
\end{itemize} 

These constraints are loose because of the large number of fibers available in the spectrograph. Even then, a significant fraction of fibers were still unassigned and therefore filled even lower priority stars and interesting, potentially extremely metal-poor (EMP, [Fe/H] $< -3.0$) MW halo stars according to Pristine.

\subsection{Data reduction}

The AAT 2DFDR\footnote{https://aat.anu.edu.au/science/software/2dfdr} package and the standard settings were used to reduce the spectra, with two small exceptions detailed in the ``Data Reduction'' Section of \citet{arentsen20} regarding the coadding of multiple spectra. The first one is that the weight attributed to each exposure is now determined by object and not by frame, while the second exception is to turn off the ADJUST\_CONTINUUM parameter that can produce unphysical CaT line shapes. 

Three examples of spectra for low (6.7), mid (14.4) and high (30.7) signal-to-noise (S/N) ratios are shown in Figure \ref{spectra_fit}. As shown by this plot, the observing run suffered from an extremely large sky contribution in each spectrum, especially in the vicinity of the third CaT line, which causes the fitting of this line to be challenging, even for high S/N, although not impossible as a minority of spectra have a prominent enough third line. Extensive testing of different sky subtraction methods, internal and external to the AAT 2DFDR software, led to the conclusion that the issue does not lie with the sky subtraction itself, which is conducted properly by the software, but by the fact that the sky contributions can be so large that even small residuals remain significant with respect to the stellar spectra. Each spectrum was therefore carefully visually inspected and discarded if its quality was too poor to obtain a proper fit of the three CaT lines. This step led to the rejection of 145 spectra, i.e. almost $50$\% of our sample. Among those, $\sim 23$ stars were high probability member candidates. A histogram of the $g_0$ magnitude of good vs. rejected spectra is shown in Figure \ref{histo_rejected_mag}, illustrating that this quality cut is made at the expense of going deeper into the RGB of Hercules.

The spectra are normalized by finding the continuum following the method of \citet{battaglia08}, i.e. through an iterative k-sigma clipping non-linear filter. The heliocentric velocities and equivalent widths (EWs) of each spectrum is then obtained using our in-house pipeline described in detail in \citet{longeard22}, that has already been extensively tested against known metallicities and velocities. Each CaT line is modeled with a Gaussian and Voigt profile and their position are found by minimizing the squared difference between a synthetic spectrum composed of three Gaussian/Voigt profiles and the observed spectrum. The EWs are calculated by integrating the best fit around each line in a 15\AA $\;$ window. This is performed with a Monte Carlo Markov Chain \citep{hastings70} algorithm with a million iterations per spectrum.

\section{Results}

We present in this Section the results of our spectroscopic analysis, both dynamical and in terms of metallicity. We start with the metallicity results since it is the most discriminative property between Hercules's stellar population and the MW's.

\subsection{Metallicity properties}

\begin{figure}
\begin{center}
\centerline{\includegraphics[width=\hsize]{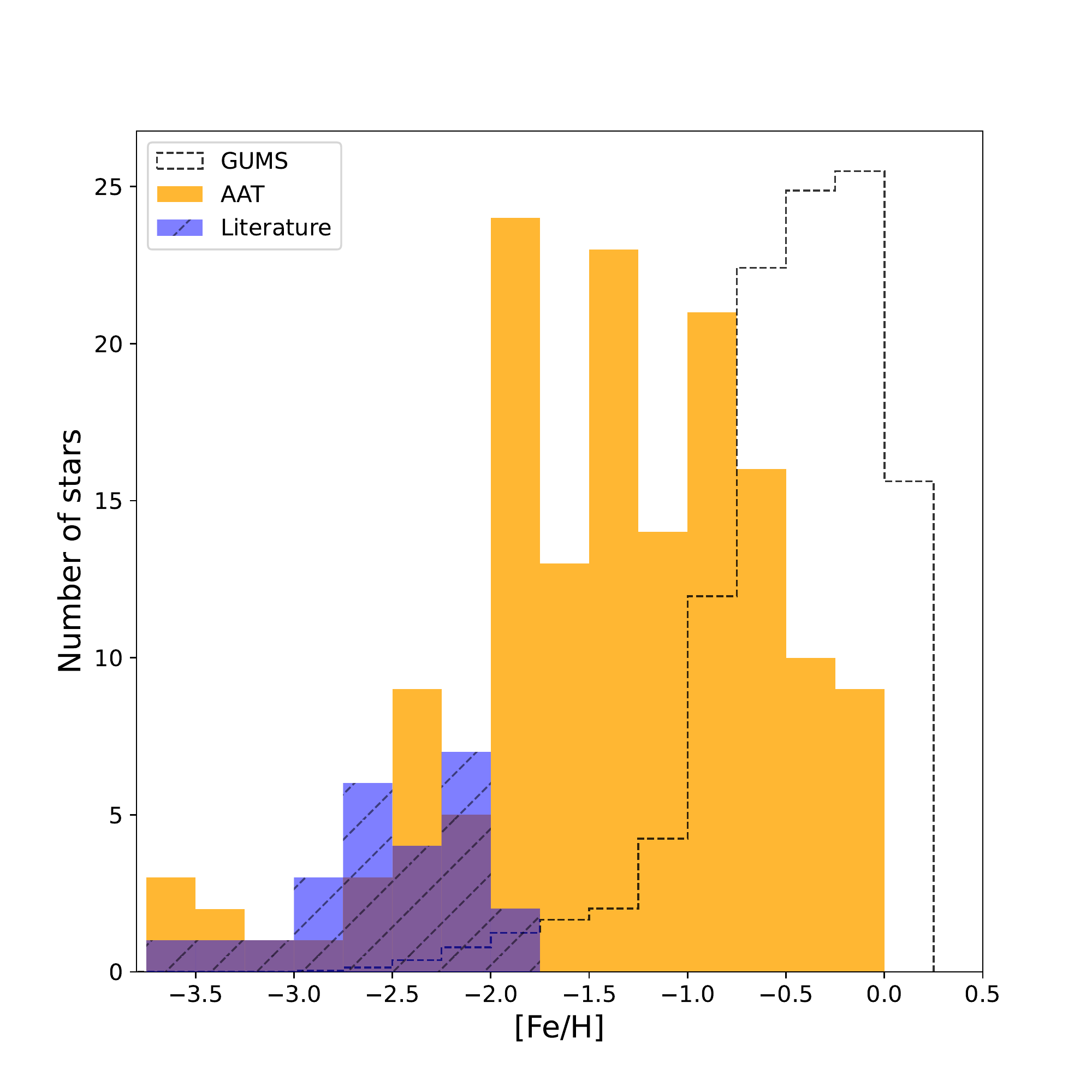}}
\caption{MDFs of the MW contamination as predicted by GUMS (dotted black), the full AAT sample (plain orange) and the literature (dashed blue). The AAT MDF is calculated with both spectroscopic and photometric metallicities from the Pristine survey. The mean metallicity of the AAT sample is naturally lower than the one of GUMS as a Pristine pre-selection on the metallicity was applied prior to observation.}
\label{MDF} 
\end{center}
\end{figure}

\begin{figure}
\begin{center}
\centerline{\includegraphics[width=\hsize]{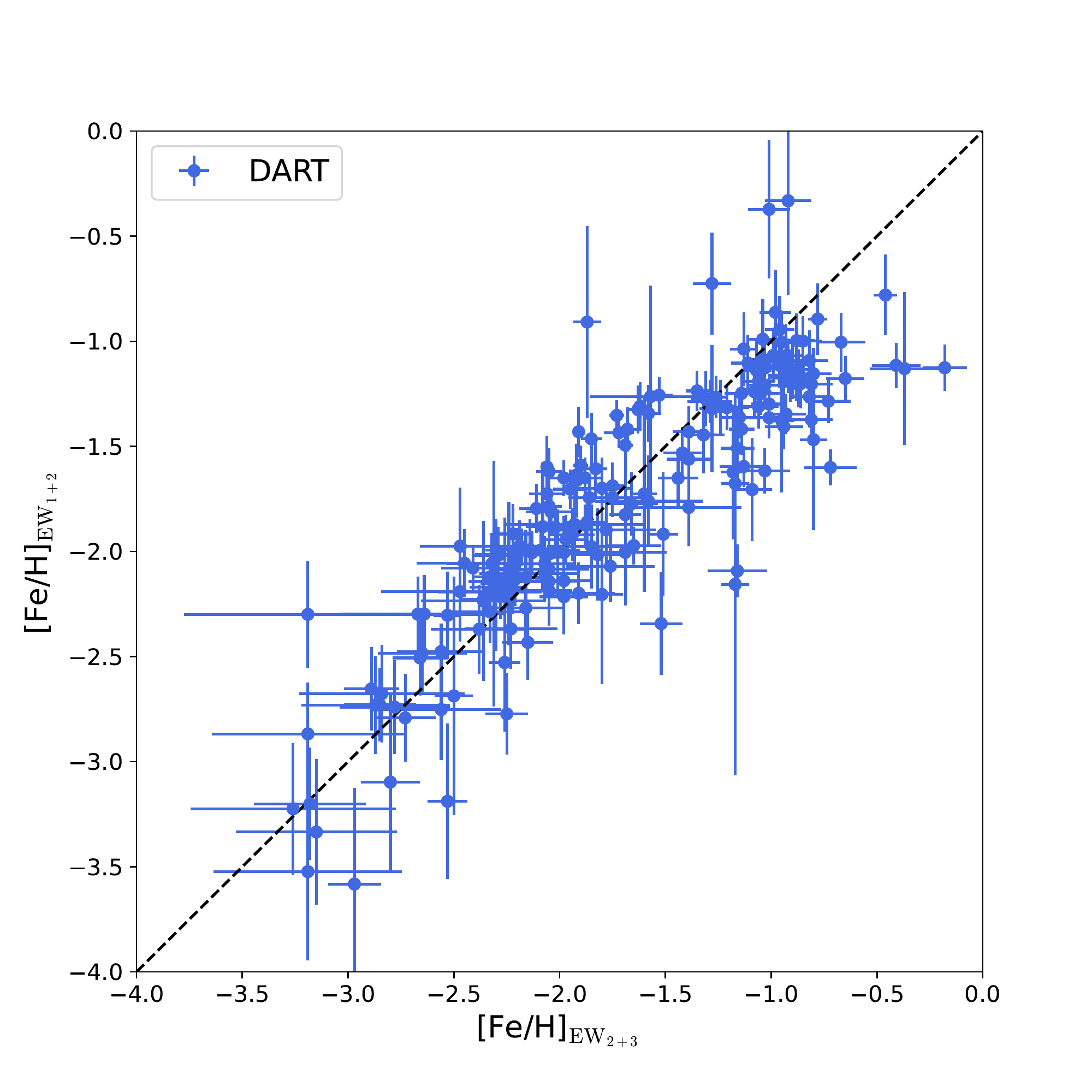}}
\caption{Comparison between metallicities obtained from the calibration of \citet{carrera13} using the second and third CaT lines (x-axis) and ours based on the first two lines (y-axis) for DART (blue). The 1:1 line is shown with the black dashed line.}
\label{spectro_vs_spectro} 
\end{center}
\end{figure}

The stellar velocity distribution of Hercules overlaps with that of the MW (see Section 3.2 for more details). Therefore, MW stars can easily be misidentified as Hercules members from a purely dynamical standpoint. An additional selection is needed. Figure \ref{MDF} shows the MDFs of the MW in the direction of Hercules as predicted by the Gaia Universe Model Snapshot \citep[GUMS]{robin12}. We superimpose the spectroscopically confirmed Hercules members from the literature. While all Hercules' stars have a spectroscopic metallicity below $-1.5$, the MW population is mostly more metal-rich, with only the tail of the [Fe/H] distribution intersecting with that of the MW. Therefore the metallicity is an appropriate way of discriminating the two populations.

However, spectroscopic metallicities derived from the CaT lines require clean spectra with a S/N of 10 at the very least. As mentioned in Section 2.2, this is not the case for most of our spectra, that not only have a lower S/N, but also are polluted with sky residuals. Moreover, the classical CaT calibrations rely on the second and third lines, since the first one typically has a lower S/N than the other two (\citealt{starkenburg10}, \citealt{carrera13}). 

A total of $29$ stars have spectra with SNR $> 10$. We can properly fit the third line of the CaT for $8$ of them. For these, we can therefore use the empirical calibration of \citet{carrera13}. Their uncertainties are derived by performing a 10,000 iterations Monte Carlo sampling on all the parameters involved in the calibration, i.e. the $V$ absolute magnitude, the distance modulus of Hercules, the EWs and the calibration coefficients.

In order to derive the CaT metallicities of the 21 remaining high SNR AAT stars, we derive a new empirical calibration based only on the two first lines based on a sample of 220 RGB stars with S/N of 10 or higher from the \textit{Dwarf galaxy Abundances and Radial-velocities Team} (DART, \citealt{tolstoy04}, \citealt{battaglia06}, \citealt{tolstoy06}, \citealt{battaglia08_b}, \citealt{battaglia11}) belonging to the Fornax, Sextans and Sculptor Dwarf Spheroidals. The same formalism as \citet{starkenburg10} and \citet{carrera13} is used, i.e.:

\begin{equation}
\mathrm{[Fe/H]}_{1+2} = a + b \times V + c \times \mathrm{EW}_{1+2} + d \times \mathrm{EW}_{1+2}^{-1.5} + e \times \mathrm{EW}_{1+2} \times V
\end{equation}

with $a$, $b$, $c$ , $d$ and $e$ the new calibration coefficients, $\mathrm{EW}_{1+2}$ the sum of the first and second CaT lines equivalent widths assuming a Voigt profile, and $V$ the $V$ absolute magnitude of the star.  Our pipeline is used to derive the EWs of the first two lines of the DART stars. The coefficients are then found by a least square minimization with respect to the ``true'' metallicities given by the DART papers, through a MCMC algorithm. The resulting coefficients are reported in Table 1. Note that these coefficients are highly correlated and cannot, as face value, be used to determine a metallicity unless one uses the full MCMC chains to draw the coefficients from. Figure \ref{spectro_vs_spectro} shows that the two calibrations are in excellent agreement. 

\begin{table}
\centering
\begin{tabular}{|llll} 

 \hline
Coefficient & Inference  \\ 
  \hline
$a$ & $-1.380 \pm 0.720 $ \\
$b$ & $0.290 \pm 0.280$ \\
$c$ & $0.024 \pm 0.200$ \\
$d$ & $-3.890 \pm 0.980$ \\
$e$ & $-0.097 \pm 0.790$ \\ 

 \hline
\end{tabular}
\caption{New calibration coefficients}
\label{table:1}
\end{table}

At this stage, all 29 AAT stars with a S/N $>10$ have a potential spectroscopic metallicity measurement. However, these CaT calibrations hold only if a star is a Hercules member, since their distance are considered when computing the spectroscopic metallicities. In order to have a metallicity estimate, irrespective of the SNR of their spectra or membership status, we also assign a Pristine metallicity estimate to the full sample. The final AAT MDF, composed of both spectroscopic metallicities (when possible) and photometric metallicities otherwise is shown in orange in Figure \ref{MDF}. The next step is to use the discriminative power of Hercules' metallicity to accurately derive its kinematic properties.

\subsection{Dynamical analysis}

\begin{figure}
\begin{center}
\centerline{\includegraphics[width=\hsize]{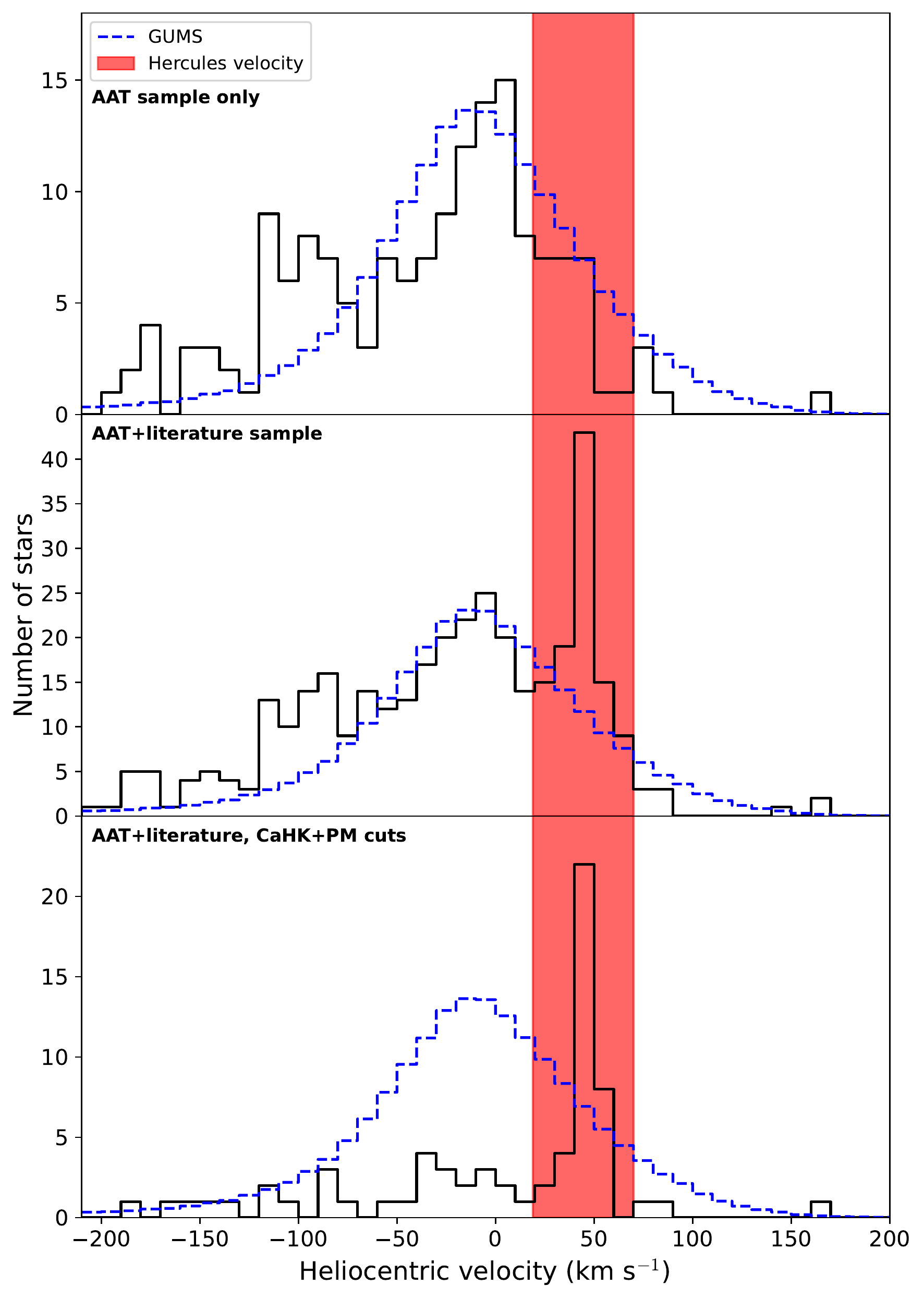}}
\caption{\textit{Top panel:} Velocity histogram for the new AAT sample. The blue dashed histogram corresponds to the GUMS velocities in the region of Hercules. The red transparent rectangle shows the $3 \; \sigma$ interval of Hercules' dynamical population. \textit{Middle panel:} Velocity histogram of the AAT sample combined with the literature. \textit{Bottom panel:} Final velocity distribution, when the AAT+literature sample is cleaned based on the photometric and Gaia criteria detailed in Section 3.2.}
\label{histo_vel} 
\end{center}
\end{figure}

\begin{figure}
\begin{center}
\centerline{\includegraphics[width=\hsize]{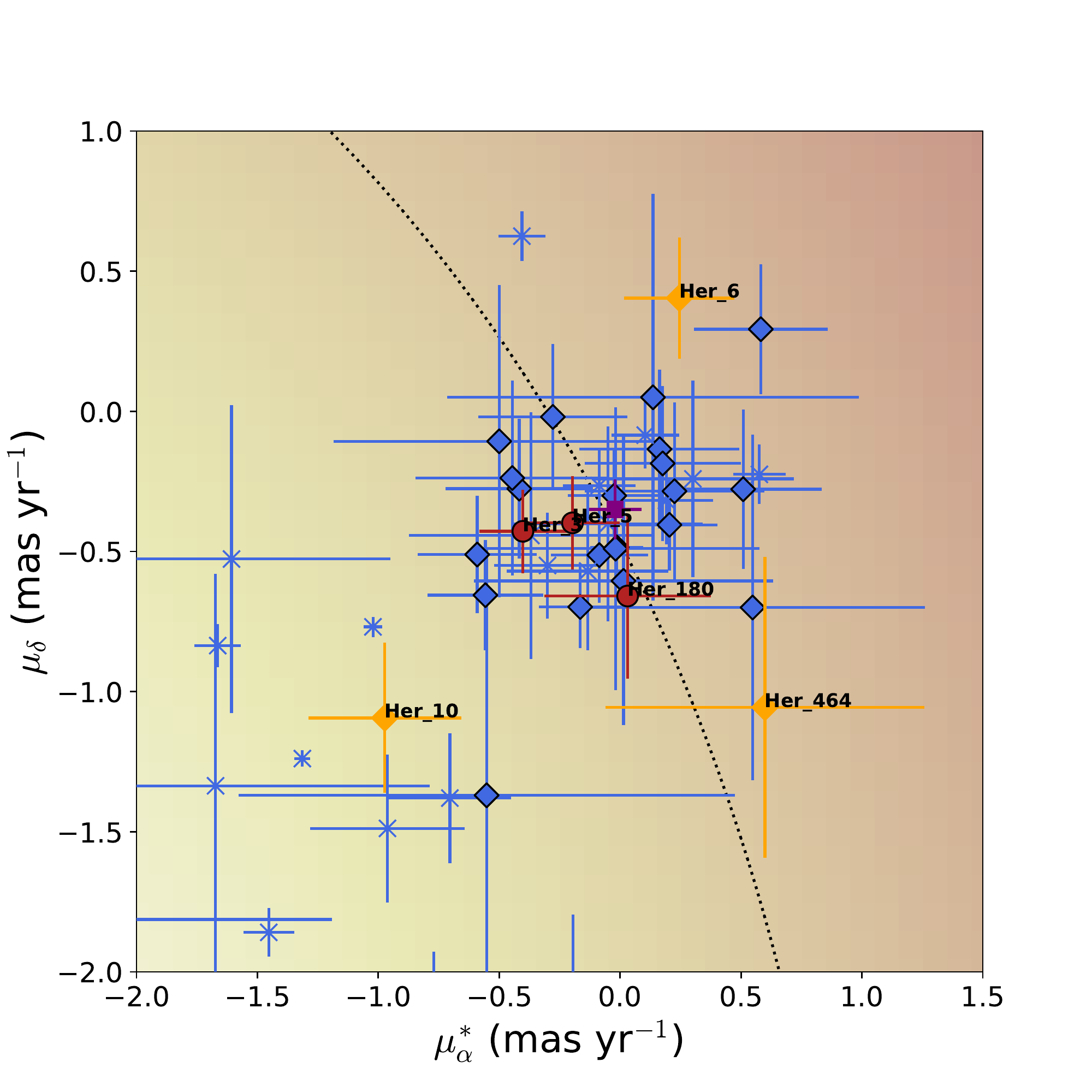}}
\caption{PM space showing the confirmed literature members as blue diamonds, literature non-members as blue crosses, and all new AAT members as red circles. Uncertain candidates are represented by orange diamonds. The large purple square shows the systemic PM of Hercules as derived in this work, compatible with the ones of \citet{battaglia22} and \citet{mcconnachie_venn20}. The underlying distribution represents the MW population in that region of the sky, assuming it can be modelled as a multivariate gaussian. The dashed black line corresponds to the 1$\sigma$ contour of this MW population.}
\label{pm_space} 
\end{center}
\end{figure}

\begin{figure}
\begin{center}
\centerline{\includegraphics[width=\hsize]{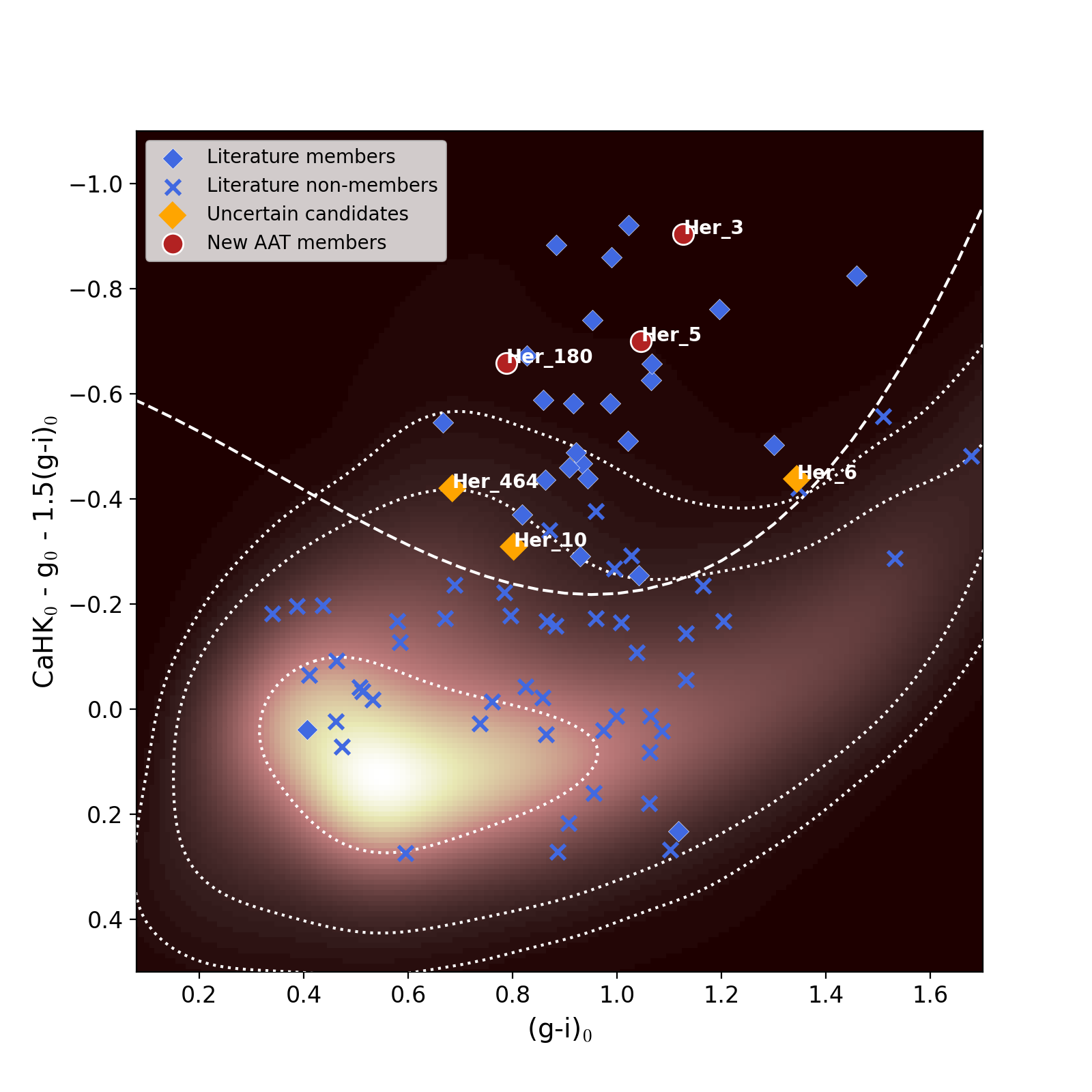}}
\caption{Pristine colour-colour space, with the ($g-i$)$_0$ temperature proxy and the $CaHK_0 - g_0 - 1.5$($g-i$)$_0$ colour on the y-axis. In this plot, [Fe/H] decreases as values on the y-axis decrease. The density plot of the MW contamination is also shown with the colour scale. The density decreases as the colour goes darker. The $1$, $2$ and $3\sigma$ contours are shown as white dotted lines. Literature members and non-members are shown as blue diamonds and crosses respectively and are only the ones with a spectroscopic metallicity and a proper motion measurement to ensure their membership status. Our new AAT members as shown as red circles, while uncertain candidates are represented as orange diamonds.}
\label{cahk_density} 
\end{center}
\end{figure}

The velocities of the new AAT sample are obtained using the pipeline described in Section 2.2. For our observational setup, \citet{li19} show that both a velocity offset and a velocity uncertainty corrections are needed. The offset is of the order of $1.1$ km s$^{-1}$, while the uncertainties are corrected using the following relation:

\begin{equation*}
\begin{aligned}
\delta_\mathrm{v} = \sqrt{(1.28\delta^\mathrm{fit}_\mathrm{v})^2 + 0.66^2},
\end{aligned}
\end{equation*}

\noindent with $\delta^{\mathrm{fit}}_\mathrm{v}$ the intrinsic velocity uncertainty derived by the CaT lines fit. The velocity measurements reported in Table 2 include these corrections.

The heliocentric velocity distributions of the new AAT sample alone and combined with the literature values are shown in Figure \ref{histo_vel}. As shown by the top panel, the case of Hercules is challenging as its velocity distribution is enclosed in the predicted MW stars' line-of-sight velocity in that region of the sky, according to GUMS. Taken alone, our AAT sample does not exhibit any velocity peak. We therefore combine the AAT dataset with previous spectroscopic studies over the years:  \citet{aden09}, \citet{martin_jin10}, \citet{deason12}, \citet{fu19} and \citet{gregory20}. The resulting sample is shown in the middle panel of Figure \ref{histo_vel}.

From this sample composed of the new AAT spectroscopy and the literature combined (368 stars in total), we reduce the potential contamination of the MW which could bias our result. To this end, we take into account the photometric (broadband and $CaHK$) and Gaia (see $\mathcal{L}^{\mathrm{Her}}_{\mathrm{PM}}$, Equation 2) information of each star. All these properties are combined in the final likelihood equation (see Equation 4) to derive the dynamical properties. For clarity, we first derive each of these likelihoods and photometric criteria separately, before combining them all to produce the final sample. 

\subsubsection{The broadband and $CaHK$ photometry information}

The initial sample is composed of $368$ stars. First, the ones with a membership probability based on their location on the Hercules CMD lower than 10\% (these probabilities are computed following the same method detailed by \citealt{longeard18}) are discarded. This leads to the exclusion of $122$ stars.

Then, a photometric metallicity cut using the CaHK photometry is applied. However, rather than using a single metallicity value as a threshold to determine the final sample, this criterion is built in the Pristine colour-colour space shown in Figure \ref{cahk_density}. In order to estimate the stellar density of the MW, we bin all stars with distances from the center of Hercules  larger than $5$r$_h$. The resulting grid is then convolved with a $0.1$ mag 2D Gaussian kernel on each axis. We then superimpose the literature spectroscopic and PM confirmed members and non-members. In this colour-colour space that is metallicity-sensitive (with metallicity decreasing as one goes upwards in the diagram), there is a clear dichotomy between the Hercules members and non-members, the former being more metal-poor than the MW stars. We therefore trace a line separating the two populations shown as the dashed white one in Figure \ref{cahk_density}. The final sample, on which the dynamical properties will be derived, is composed solely of stars located in the region above that line. In doing so, we take the risk of missing some metal-rich Hercules members located in the discarded area. However, this risk should be extremely low as most of our targets are located in the outskirts of the galaxy which are expected to have a lower metallicity than that of the main body. This is true even for UFDs, as demonstrated recently by \citet{chiti21} for Tucana~II and by \citet{longeard22} for Bootes~I. This cut further excludes 157 stars.

\subsubsection{The Gaia information}

We now have a sample composed of $89$ stars. We now apply an effective temperature and parallax cut: every star with T$_\mathrm{eff} > 5800$ K and $\mathrm{Plx}/\mathrm{e\_plx} > 2.0$ are discarded, with T$_\mathrm{eff}$ the effective temperature measured by Gaia DR3, Plx the parallax and $\mathrm{e\_plx}$ its corresponding uncertainty. These criteria discard $13$ stars. 

The PM information is then folded in by estimating the local MW contamination in PM space assuming a 2D Gaussian mixture models to the PM distribution shown in Figure \ref{pm_space}. This step corresponds to the likelihood:

\begin{equation}
\begin{aligned}
\mathcal{L}^{\mathrm{Her}}_{\mathrm{PM}}(\langle \mu_{\alpha}^{*} \rangle, \langle \mu_{\delta}\rangle, \sigma_{\alpha}, \sigma_{\delta}, c | \mu^{*}_{\alpha,k}, \mu_{\delta,k}, d\mu^{*}_{\alpha,k}, d\mu_{\delta,k}) = \\
\mathcal{G}_{2D}(\mu^{*}_{\alpha,k},\mu_{\delta,k},d\mu^{*}_{\alpha,k}, d\mu_{\delta,k} | \langle \mu_{\alpha}^{*} \rangle, \langle \mu_{\delta}\rangle, \sigma_{\alpha}, \sigma_{\delta}))
\end{aligned}
\end{equation}

\noindent with $\langle \mu^{*} \rangle$ and $\sigma$ the systemic proper motion and proper motion dispersion, $c$ the correlation and $\mu^{*}_{k}$ the individual PM measurement. The ``$\mathcal{G}_{2D}$'' notation stands for a 2D Gaussian function. This equation is only shown as a functional form and can also be applied to the MW. This step is then folded in the final likelihood Equation 4. It is not a dichotomic cut, i.e. stars are not discarded of the sample altogether based on their PM measurements, but their PM membership probability following Eq. 2 will intervene in the kinematic properties derivation, through Eq. 4 detailed later. For stars without a PM measurement, we set their PM to be 0 mas yr$^{-1}$ and their PM uncertainty to be $10^{6}$ mas yr$^{-1}$.

From the photometric cuts and the PM information, a total of 11 literature members are found not to be members of Hercules. Their properties are detailed in Table 6, with the last column indicating the main information used to make the decision. In particular, it is interesting to note that the two most metal-rich literature members belong to this misidentified member sample, since their PM is largely discrepant from that of Hercules.

\subsubsection{The kinematic analysis}

\begin{figure}
\begin{center}
\centerline{\includegraphics[width=\hsize]{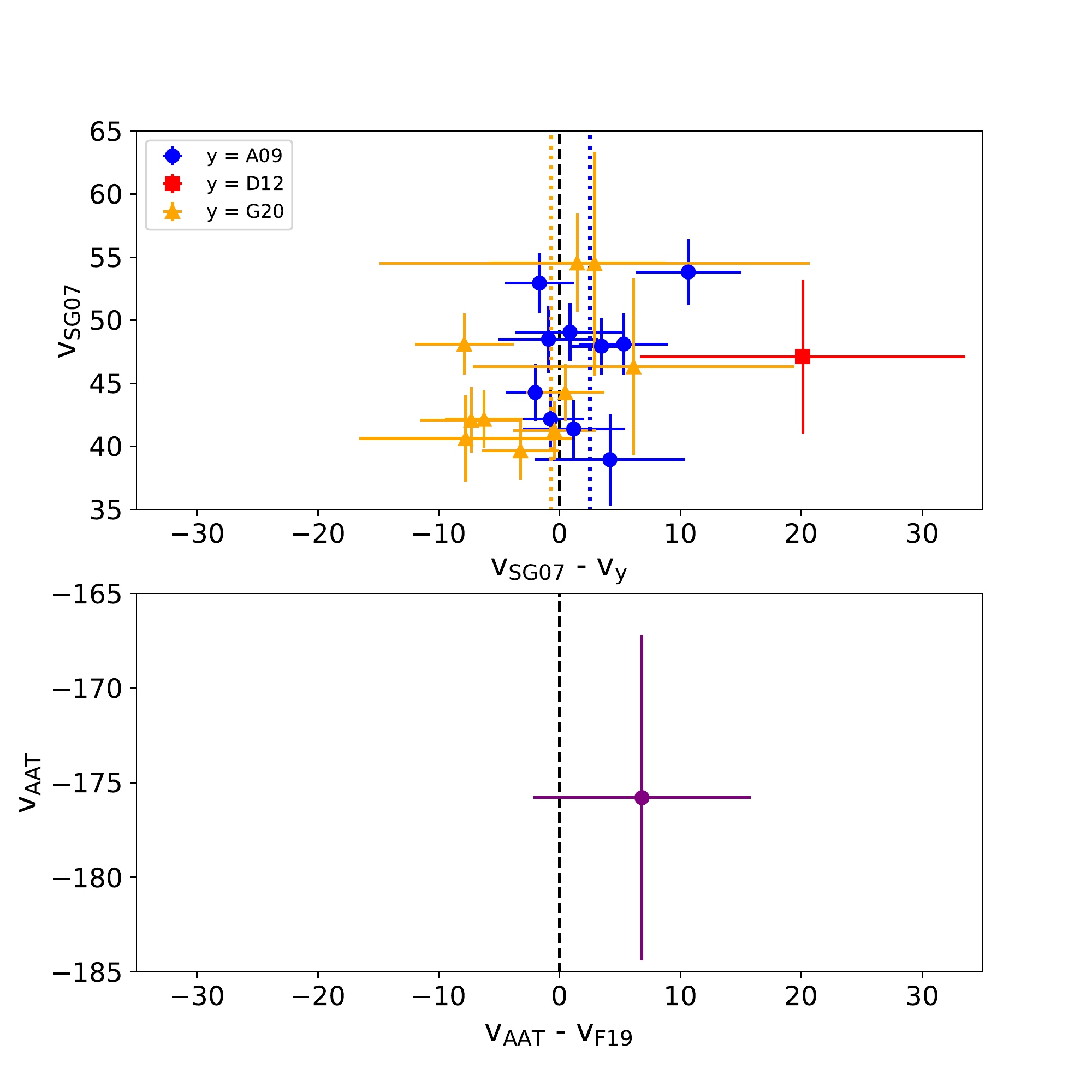}}
\caption{\textit{Upper panel: }Heliocentric velocity differences between the sample of \citet{simon_geha07} and the ones of \citealt{aden09} (A09, blue circles), \citealt{deason12} (D12, red squares) and \citealt{gregory20} (G20, orange triangles). \textit{Lower panel: }Heliocentric velocity difference of a star in common between the new AAT sample and the one of \citet{fu19}. The central dashed black line shows the identity. Differences larger than 20 km s$^{-1}$ have been excluded as they most likely reflect poor fitting on one or both side(s).}
\label{diff_vel} 
\end{center}
\end{figure}

\begin{figure*}
\begin{center}
\centerline{\includegraphics[width=\hsize]{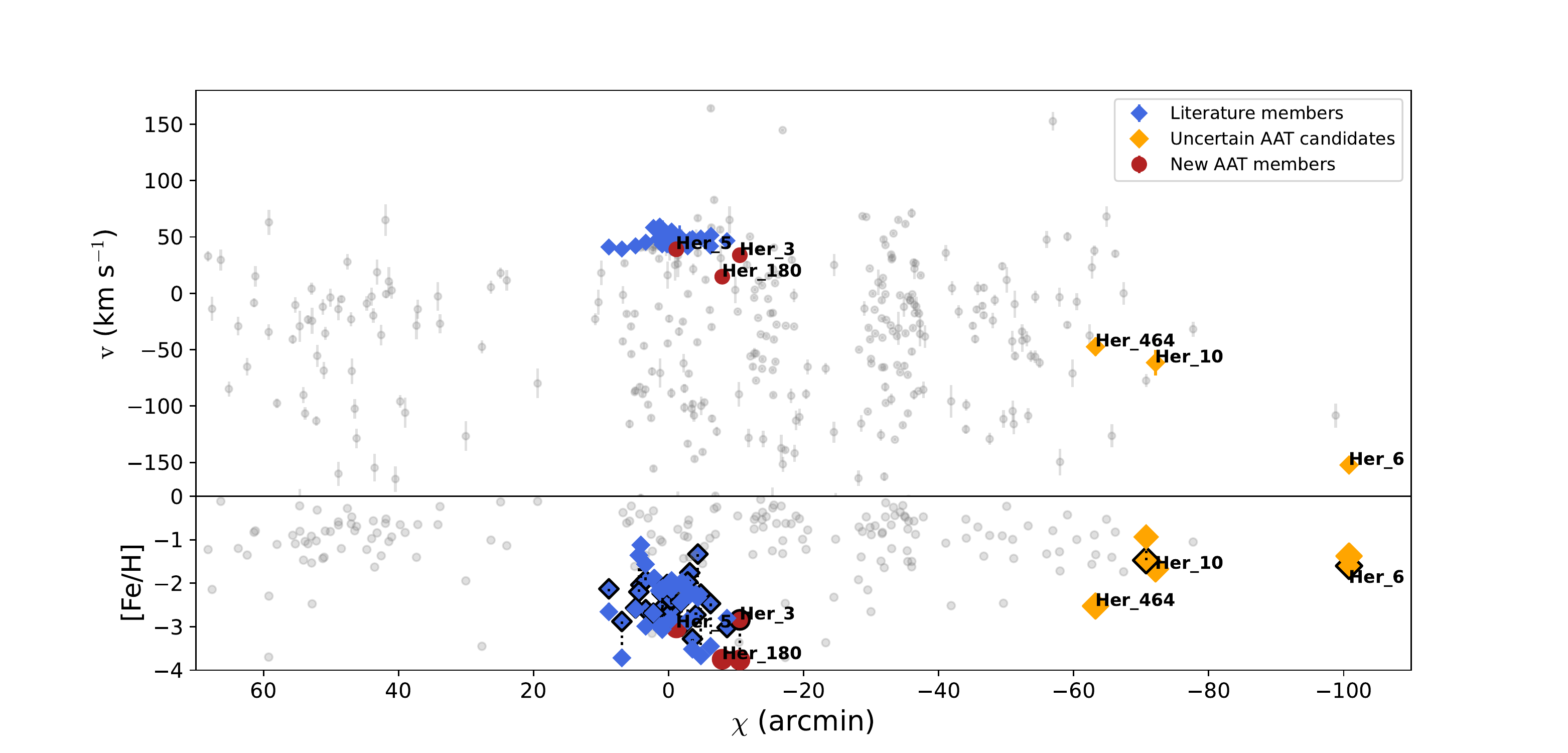}}
\caption{\textit{Top panel:} Phase-space distribution of the AAT+literature non-members (grey), literature members (blue diamonds), AAT uncertain candidates (orange diamonds) and AAT new confirmed members (red circles). $\chi$ is defined as a distance respective to a given position angle (see section 3.2.3). \textit{Bottom panel:} Position vs. metallicity. When both photometric and spectroscopic metallicities are available for the same star, they are linked by a vertical dashed black line, with the black-contoured symbols indicating the spectroscopic one.}
\label{v_vs_r} 
\end{center}
\end{figure*}

At this stage, the remaining sample is composed of $76$ stars, both from the literature and the new AAT spectroscopic sample. 

The final step before inferring Hercules' new dynamical properties is to properly handle the potential velocity offsets between the different spectroscopic datasets. They are indeed observed at different times with different setups, which can introduce velocity offsets between different datasets that need to be corrected for. Ideally, using the formalism of \citet{minor19} that proposes to add as many unknown offset parameters to infer as there are different spectroscopic setups to find the systematic differences between them would be sufficient. However, this method fails for small datasets and/or when the UFD's population is blended with that of the MW. For \citet{aden09}, \citet{gregory20} and \citet{fu19}, a cross-match is therefore performed between each dataset and the reference one of \citet{simon_geha07}. The offset is then derived by measuring the mean velocity difference between all stars in common. The uncertainty on each offset is taken as the standard deviation of the velocity differences. 

Finally, for \citet{deason12}, the same method is applied but in two steps: the offset between this catalog and the one of \citet{aden09} is found. Then, we apply a second offset between the latter and the \citet{simon_geha07} dataset. The reason for this intermediate step is that the cross-match between the datasets of \citet{deason12} and \citet{aden09} yields more stars in common than directly cross-matching with \citet{simon_geha07}, therefore constraining better the offset.

Finally, for our AAT sample, the offset is also found in two steps. First, the one from \citet{longeard22} between the setup used in this work and the one of the reference sample, i.e. a Keck/DEIMOS setup in the red, of $7.2 \pm 1.6$ km s$^{-1}$, used by \citet{martin07}, is considered. Then, using the cross-match of the datasets of \citet{simon_geha07} and \citet{martin07} for three other dwarf galaxies, we find a systematic of $2.7 \pm 1.8 $ km s$^{-1}$, which is added to the one of \citet{longeard22} to give the final offset between these two setups. The velocity difference of the star in common between the reference dataset and the AAT sample, of $\sim 8$ km s$^{-1}$, supports this choice. Figure \ref{diff_vel} presents the velocity differences between the different spectroscopic samples. \\

The dynamical properties of Hercules are derived following the formalism of \citet{martin_jin10} combined with the likelihoods described in Equations (2) and (3):

\begin{equation}
\begin{aligned}
\mathcal{L}(\langle \mathrm{v}_{\mathrm{Her}} \rangle, \langle \mathrm{v}_{\mathrm{MW}} \rangle, \sigma^\mathrm{Her}_{\mathrm{v}}, \sigma^\mathrm{MW}_{\mathrm{v}}, d\mathrm{v}/d\chi, \theta | \mathrm{v}_{\mathrm{r},k}, \delta_{\mathrm{v},k}) = \\
\prod_k \;  \bigg[ \eta_\mathrm{Her} (\frac{1}{\sqrt{2 \pi \sigma}}) \times \mathrm{exp}(\frac{1}{2}\Delta_\mathrm{v}/\sigma^{2}) \mathcal{L}^{\mathrm{Her}}_{\mathrm{PM}}+ \\
(1 - \eta_\mathrm{Her}) \mathcal{G}(\mathrm{v}_{\mathrm{r},k}, \delta_{\mathrm{v},k},\langle \mathrm{v}_{\mathrm{MW}} \rangle, \sigma^\mathrm{MW}_{\mathrm{v}}) \mathcal{L}^{\mathrm{MW}}_{\mathrm{PM}} \bigg],
\end{aligned}
\end{equation}

\noindent We define $\Delta_\mathrm{v}$ such that $\Delta_\mathrm{v} = \mathrm{v}_{\mathrm{r},k} - y \times d\mathrm{v}/d\chi + \langle \mathrm{v}_{\mathrm{Her}} \rangle$ with d$\mathrm{v}$/d$\chi$ the systemic heliocentric velocity gradient. $y$ is the angular distance computed such that $y_k = X_k\sin{\theta} + Y_k\cos{\theta}$ and $\theta$ the direction of the velocity gradient. We also define $\sigma = \sqrt{(\sigma^\mathrm{Her}_\mathrm{v} + \delta_\mathrm{v}^{2})}$. The velocity gradient defined in such a model corresponds to a linear, monotonic velocity change along the dwarf. We adopt the same convention as previous dynamical analyses of Hercules, i.e. that a positive velocity gradient corresponds to an increase of the velocity towards decreasing RA. A 2,000,000 iterations MCMC is performed, and the results are shown in Figure \ref{pdfs_vel_grad}. Based on this analysis, a kinematic membership probability can be computed for all stars. Combined with their metallicity properties, they lead to the discovery of $3$ new Hercules members, including one at $\sim 9.5$R$_h$ ($\sim 2.1$ kpc) of the satellite, computed as the elliptical distance to Hercules' centroid according to the structural parameters of \citet{munoz18}. The others lie at $\sim 1.6$R$_h$ ($\sim 335$ pc) and $\sim 0.5$R$_h$ ($\sim 97$ pc). We also report $3$ ``uncertain candidates'', i.e. stars marginally compatible with Hercules' dynamical properties, but with an unconvincing membership status discussed in Section 4. The locations of these $6$ stars are shown in Fig. \ref{field}, \ref{pm_space}, \ref{cahk_density} and \ref{v_vs_r}.

We find a systemic velocity of $\langle \mathrm{v} \rangle = 45.7^{+2.3}_{-3.7}$ km s$^{-1}$, an intrinsic velocity dispersion of $8.0^{+1.4}_{-2.0}$ km s$^{-1}$, at odds with measurements from the literature ($5.1 \pm 0.9$ km s$^{-1}$). Finally, no significant velocity gradient is detected with an inference of $0.1 _{-0.2}^{+0.4}$ km s$^{-1}$ arcmin$^{-1}$, i.e. $1.6^{+10.0}_{-3.8}$ km s$^{-1}$ kpc$^{-1}$. The posterior PDF for the velocity gradient is shown as the dotted blue line in Figure \ref{pdfs_vel_grad}.

\section{Membership assessment}

Figure \ref{v_vs_r} shows the distribution of our AAT sample superimposed with literature members. Two different groups can be identified in the new spectroscopy, those being the new members including all criteria detailed in Section 3.2, and the uncertain candidates. The membership of each star will be discussed individually in what follows.

\subsection{The new members Her~3, Her~5 and Her~180}
Their properties strongly favour them to be bona-fide members of Hercules. First of all, they lie on the RGB of the UFD (Figure 1). Their PM and heliocentric velocity membership are also extremely high, i.e. more than $90$\% in all cases. However, as detailed above, these properties are entangled with the MW. The most compelling evidence of their membership lie in the Pristine photometry presented in Figure \ref{cahk_density}, where they lie in a very metal-poor region of the diagram.

Ideally, this would be confirmed by a spectroscopic derivation of their metallicities. While the S/N of Her~5 and Her~180 spectra are slightly too low to yield a reliable metallicity estimate, it is possible for Her 3 (S/N $\sim 12.7$) using the calibration presented in this work based on the first two CaT lines. Its spectroscopic metallicity of [Fe/H] $= -3.0 \pm 0.5$ places it well within the MDF of Hercules and in the tail of the MW's (Figure \ref{MDF}).

The only element of caution regarding Her~5 is its spatial location, as Figure \ref{field} shows that the star is not located along the major axis of the very elongated UFD while still located at a very large distance ($\sim 9.5$r$_h$). This fact is however mitigated by the finding of \citet{garling18} and one of their new RR lyrae located well beyond Hercules' tidal radius while also being far off its major axis. Her~3 and Her~180 positions are aligned with the major axis of Hercules.

\subsection{The uncertain candidates}

We also find $3$ uncertain candidates shown as the orange diamonds in Figure \ref{v_vs_r}. These stars have a proper motion compatible with that of Hercules but lie at the edge of the CaHK selection detailed in section 3.2.1, where contamination from the MW is still not unlikely.

Based on their CaHK colour-colour diagram locations, which place them at the edge of the members/non-members region defined in Figure \ref{cahk_density}, Her~464 and Her~10 should be significantly more metal-rich than our $3$ new members. However, this is absolutely not reflected in their locations in the CMD of Hercules. Only Her~6 has a coherent CMD location since it is redder than the favoured Hercules isochrone. One of these also have a spectroscopic metallicity measurement (which is only valid if it is considered to be at Hercules' distance) placing it at the metal-rich rail of Hercules' MDF. Such a metallicity would be unlikely for distant members since the metal-rich population is supposed to be more centrally concentrated even in UFDs (\citealt{chiti21}, \citealt{longeard22}).
Furthermore, while Her~464's PM uncertainties allow for it to be perfectly compatible with the systemic proper motion of Hercules, the other two stars PMs are slightly less convincing. 

Finally, Figure \ref{v_vs_r} show a suspicious spatial gap between $-20$ and $-60$ arcmin, i.e. between our confirmed members and the uncertain candidates. The finding of a member in this gap would have brought credit to at least one of the candidates to be members, but even pushing down our kinematic membership probability down to a threshold of $1$\% (while keeping the CaHK selection) yields no potential members in that area.

For all these reasons, we favour these $3$ stars to not be Hercules members. This decision has an impact of the resulting velocity gradient as shown in the next section.

\section{On the velocity gradient of Hercules}

\begin{figure}
\begin{center}
\centerline{\includegraphics[width=\hsize]{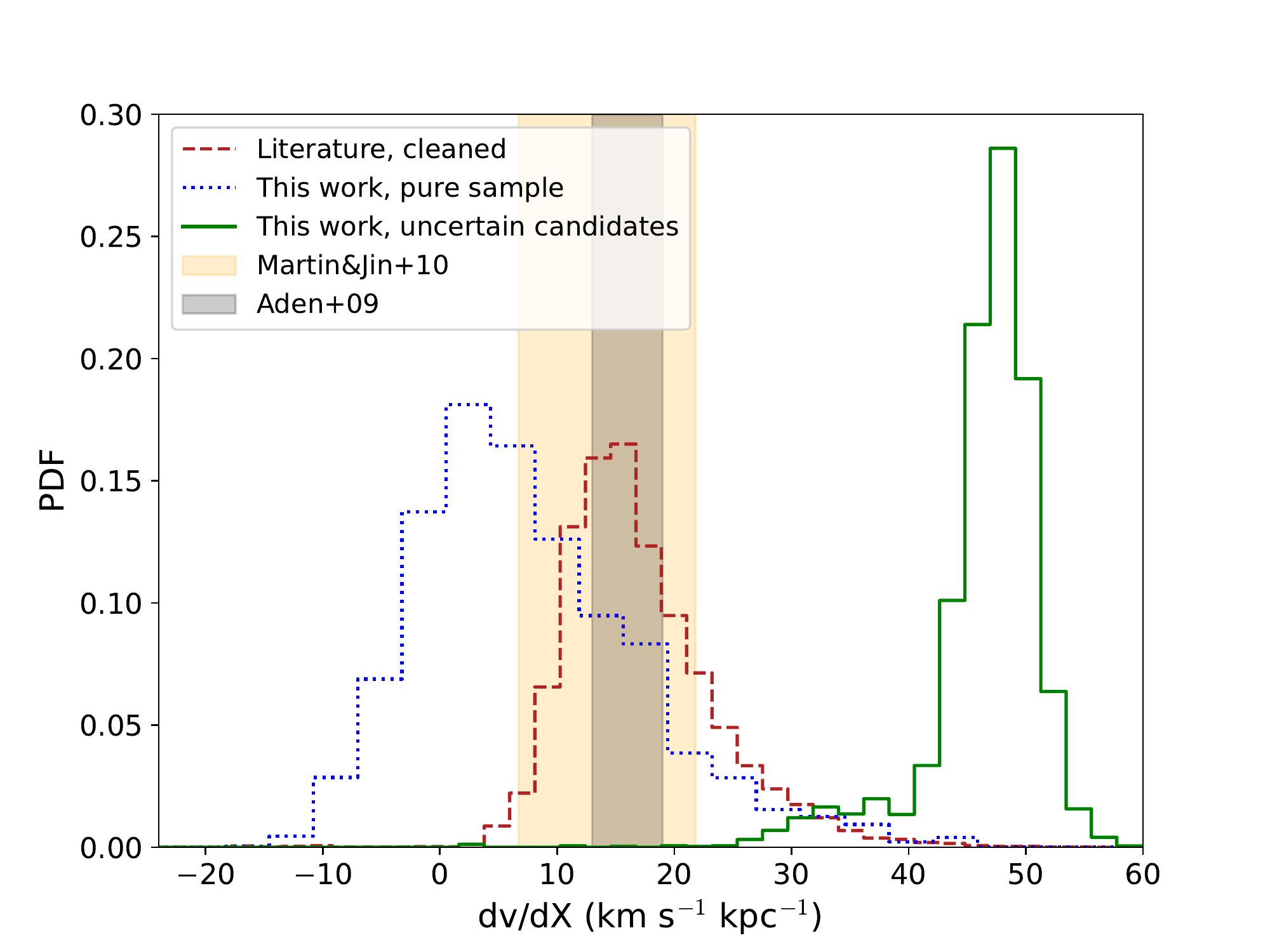}}
\caption{PDFs of the velocity gradient of Hercules in the three cases detailed in section 5.}
\label{pdfs_vel_grad} 
\end{center}
\end{figure}

Our measurement excludes the observed gradient of \citet{aden09} ($16 \pm 3$ km s$^{-1}$ kpc$^{-1}$), but does not discard the theoretical ones of \citet{fu19} ($0.6$ km s$^{-1}$ kpc$^{-1}$) or \citet{kupper17} ($4.9$ km s$^{-1}$ kpc$^{-1}$). This result is driven by the rejection of $3$ stars classified as uncertain candidates. To better understand the impact of this choice, the dynamical modelling analysis detailed in the last section is performed in three different cases:

\begin{itemize}
\item Only spectroscopy from the literature, cleaned with the Gaia selection (when available, section 3.2.1)  but without the CaHK selection (section 3.2.2) (dashed red line, see Figure \ref{pdfs_vel_grad})
\item The final spectroscopic sample detailed in section 3.2 including the $3$ uncertain candidates (solid green line).
\item The final spectroscopic sample detailed in section 3.2 without the $3$ uncertain candidates (dotted blue line).
\end{itemize}

The results are shown in Figure \ref{pdfs_vel_grad}. It confirms that our analysis is consistent with previous studies as using only the literature spectroscopy combined Gaia yields a similar gradient as the ones found by \citet{aden09} and \citet{martin_jin10}. However, once the AAT data and the metallicity-sensitive, CaHK photometry are introduced, two cases are possible. If the $3$ uncertain stars at large distances are considered as members, then a significant velocity gradient is found ($45.9^{+2.7}_{-2.6}$ km s$^{-1}$ kpc$^{-1}$). However, if they are considered as MW contaminants, no statistically significant gradient is found. Given the large doubt casted on their membership as detailed in Section 4, this hypothesis is favoured.


\section{Discussion and conclusion}

We present in this work new medium resolution spectroscopy of the faint and extremely elongated UFD Hercules with the AAT and its 2dF spectrograph. A total of 175 spectra with SNR $\geq 3$ of high and low-probability potential members were analysed as far as $\sim 13$ r$_h$ of the UFD. This new sample is then combined with all available literature datasets. The CMD location, PM, and metallicity-sensitive, CaHK magnitude of each star are used to clean the spectroscopy from obvious contaminants. While the velocities and PMs of Hercules' stellar population are useful but do not give enough discriminating power with respect to MW halo stars, the Pristine photometry is an ideal tool to separate the two populations (Figure \ref{cahk_density}). This leads to the rejection of 11 literature stars previously misclassified as Hercules members. Among those are the two most metal-rich stars in Hercules according to the literature, at [Fe/H] $\sim -1.5$ and [Fe/H] $\sim -1.3$ respectively (see Table 3), rejected because of their proper motion. This shows that Hercules' star formation history is most likely shorter than previously thought. 

Furthermore, we report the discovery of $3$ new member stars, including one located at $9.5$r$_h$ of the system, as well as $3$ uncertain candidates likely to be MW contaminants. With this new sample, we find an inflated velocity dispersion ($8.0^{+1.4}_{-2.0}$ km s$^{-1}$ km s$^{-1}$) and no statistically significant velocity gradient ($1.6^{+10.0}_{-3.8}$ km s$^{-1}$ kpc$^{-1}$). In the unlikely event where at least one of the $3$ uncertain stars is a member, the velocity gradient would be larger than any observed measurement or theoretical predictions for Hercules until now ($45.9^{+2.7}_{-2.6}$ km s$^{-1}$ kpc$^{-1}$), which would point towards significant tidal interactions with the MW.

We simulated the CMD of a system with a similar luminosity as that of Hercules based on its best-fitting isochrone. Comparing the number of simulated RGB stars with the bona-fide, observed RGB members of Hercules shows that they should all have been identified. Therefore, if a significant amount of new RGB stars are discovered in the future, it would indeed mean that Hercules used to be more luminous, and therefore more massive.

Hercules has been over the years the subject of speculations as to its degree of tidal disruption, triggered first by its extremely elongated shape. The detection of an actual velocity gradient has been debated (\citealt{deason12}, \citealt{gregory20}), and N-body dynamical simulations of its orbit show that the satellite, if tidally disrupting, should exhibit a positive velocity gradient along its major axis, although its theoretical expected magnitude is also not clear (\citealt{kupper17}, \citealt{fu19}). Considering only the new AAT convincing Hercules members, this work does not show any evidence of a significant velocity gradient in Hercules, nor that it is part of a tidal stream. Finally, we also confirm the existence of an extremely large galactocentric star in the system, and more generally detected in some UFDs.

\section*{Data availability}

The data underlying this article are available in the article.

\section*{Acknowledgments}


This work has been carried out thanks to the support of the Swiss National Science Foundation.

Based on observations obtained with MegaPrime/MegaCam, a joint project of CFHT and CEA/DAPNIA, at the Canada-France-Hawaii Telescope (CFHT) which is operated by the National Research Council (NRC) of Canada, the Institut National des Science de l'Univers of the Centre National de la Recherche Scientifique (CNRS) of France, and the University of Hawaii.

The authors thank the International Space Science Institute, Bern, Switzerland for providing financial support and meeting facilities to the international team Pristine.

NFM gratefully acknowledge support from the French National Research Agency (ANR) funded project ``Pristine'' (ANR-18-CE31-0017) along with funding from the European Research Council (ERC) under the European Unions Horizon 2020 research and innovation programme (grant agreement No. 834148).

This work has made use of data from the European Space Agency (ESA) mission Gaia (https://www.cosmos. esa.int/gaia), processed by the Gaia Data Processing and Analysis Consortium (DPAC, https://www.cosmos.esa. int/web/gaia/dpac/consortium). Funding for the DPAC has been provided by national institutions, in particular the institutions participating in the Gaia Multilateral Agreement.

The Pan-STARRS1 Surveys (PS1) and the PS1 public science archive have been made possible through contributions by the Institute for Astronomy, the University of Hawaii, the Pan-STARRS Project Office, the Max-Planck Society and its participating institutes, the Max Planck Institute for Astronomy, Heidelberg and the Max Planck Institute for Extraterrestrial Physics, Garching, The Johns Hopkins University, Durham University, the University of Edinburgh, the Queen's University Belfast, the Harvard-Smithsonian Center for Astrophysics, the Las Cumbres Observatory Global Telescope Network Incorporated, the National Central University of Taiwan, the Space Telescope Science Institute, the National Aeronautics and Space Administration under Grant No. NNX08AR22G issued through the Planetary Science Division of the NASA Science Mission Directorate, the National Science Foundation Grant No. AST-1238877, the University of Maryland, Eotvos Lorand University (ELTE), the Los Alamos National Laboratory, and the Gordon and Betty Moore Foundation.

This project has received funding from the European Union's Horizon 2020 research and innovation programme under grant agreement No 730890. This material reflects only the authors views and the Commission is not liable for any use that may be made of the information contained therein.

Based (in part) on data acquired at the Anglo-Australian Telescope, under OPTICON 2022A/041. We acknowledge the traditional custodians of the land on which the AAT stands, the Gamilaraay people, and pay our respects to elders past and present.

We thank Ángel Rafael Lopez Sanchez who observed the AAT data used in this work.

We also thank Anke Arentsen for providing us the tools to more easily perform the data reduction.

\newpage

\begin{table*}

\caption{Properties of the new AAT spectroscopic sample. The individual spectroscopic metallicities for our calibration (i.e. using CaT lines 1+2) are reported for member stars with S/N $\geq 10$. Confirmed new members are denoted with "Y" in the member column, while uncertain candidates according to the definition in section 3.2.3 are denoted with ``?''.
\label{table1}}

\setlength{\tabcolsep}{2.5pt}
\renewcommand{\arraystretch}{0.3}
\begin{sideways}
\begin{tabular}{cccccccccccccc}
\hline
RA (deg) & DEC (deg) & $g_0$ & $i_0$ & $CaHK_0$ & $\mathrm{v}_{r} (\kms)$ & $\mu_{\alpha}^{*}$ (mas.yr$^{-1}$) & $\mu_{\delta}$ (mas.yr$^{-1}$) &  S/N & [Fe/H]$^\mathrm{spectro}_\mathrm{1+2}$ & [Fe/H]$_\mathrm{Pristine}$ & Member\\

\hline

246.87308 & 13.56735 & 19.47 $\pm$ 0.02 & 18.36 $\pm$ 0.01 & 21.05 $\pm$ 0.08 & $-$80.1 $\pm$ 4.2 & $-$0.18 $\pm$ 0.18 & $-$0.27 $\pm$ 0.15 & 14.4 & --- & $-$0.57 &  N  \\ \\ 
246.60181 & 12.40716 & 20.02 $\pm$ 0.02 & 19.22 $\pm$ 0.02 & 20.91 $\pm$ 0.08 & $-$51.7 $\pm$ 11.5 & $-$0.97 $\pm$ 0.32 & $-$1.09 $\pm$ 0.27 & 3.0 & --- & $-$1.67 &  ?  \\ \\ 
248.81762 & 12.76141 & 19.29 $\pm$ 0.02 & 18.33 $\pm$ 0.01 & 20.51 $\pm$ 0.06 & $-$87.7 $\pm$ 4.2 & $-$3.86 $\pm$ 0.14 & $-$6.46 $\pm$ 0.12 & 13.1 & --- & $-$1.11 &  N  \\ \\ 
248.78361 & 13.04866 & 19.93 $\pm$ 0.02 & 18.75 $\pm$ 0.02 & 21.47 $\pm$ 0.12 & 24.9 $\pm$ 9.1 & $-$1.54 $\pm$ 0.22 & $-$4.12 $\pm$ 0.18 & 5.3 & --- & $-$0.80 &  N  \\ \\ 
247.55662 & 13.38471 & 19.76 $\pm$ 0.02 & 18.49 $\pm$ 0.01 & 21.32 $\pm$ 0.10 & 43.2 $\pm$ 5.8 & 2.15 $\pm$ 0.16 & $-$13.18 $\pm$ 0.13 & 12.4 & --- & $-$1.20 &  N  \\ \\ 
248.80239 & 13.00075 & 18.93 $\pm$ 0.01 & 17.77 $\pm$ 0.01 & 20.42 $\pm$ 0.07 & 1.3 $\pm$ 4.0 & $-$0.08 $\pm$ 0.10 & 0.15 $\pm$ 0.09 & 11.4 & --- & $-$0.83 &  N  \\ \\ 
246.69826 & 12.36966 & 20.32 $\pm$ 0.03 & 19.63 $\pm$ 0.02 & 21.09 $\pm$ 0.10 & 9.8 $\pm$ 9.9 & $-$0.61 $\pm$ 0.47 & $-$1.67 $\pm$ 0.39 & 4.4 & --- & $-$1.74 &  N  \\ \\ 
247.50016 & 12.52366 & 20.48 $\pm$ 0.03 & 19.80 $\pm$ 0.03 & 21.41 $\pm$ 0.12 & $-$197.4 $\pm$ 10.1 & $-$0.62 $\pm$ 0.42 & $-$1.60 $\pm$ 0.35 & 3.7 & --- & $-$0.99 &  N  \\ \\ 
247.50492 & 13.03883 & 20.30 $\pm$ 0.03 & 19.68 $\pm$ 0.03 & 20.75 $\pm$ 0.07 & $-$79.8 $\pm$ 10.8 & $-$1.50 $\pm$ 0.41 & $-$0.95 $\pm$ 0.34 & 6.2 & --- & $-$3.36 &  N  \\ \\ 
247.65178 & 12.9711 & 20.35 $\pm$ 0.03 & 19.24 $\pm$ 0.02 & 22.16 $\pm$ 0.20 & $-$92.5 $\pm$ 7.6 & $-$0.56 $\pm$ 0.32 & $-$0.61 $\pm$ 0.27 & 5.2 & --- & $---$ &  N  \\ \\ 
246.76058 & 13.02012 & 20.32 $\pm$ 0.03 & 19.66 $\pm$ 0.03 & 21.28 $\pm$ 0.11 & $-$24.1 $\pm$ 6.6 & $-$0.12 $\pm$ 0.58 & 0.94 $\pm$ 0.59 & 3.8 & --- & $-$0.68 &  N  \\ \\ 
246.09885 & 12.4867 & 20.36 $\pm$ 0.03 & 19.66 $\pm$ 0.02 & 21.30 $\pm$ 0.10 & $-$98.6 $\pm$ 10.7 & $-$1.39 $\pm$ 0.47 & $-$1.76 $\pm$ 0.44 & 3.9 & --- & $-$0.97 &  N  \\ \\ 
247.63172 & 12.77975 & 20.40 $\pm$ 0.03 & 19.61 $\pm$ 0.02 & 20.92 $\pm$ 0.09 & 24.5 $\pm$ 6.5 & 0.03 $\pm$ 0.34 & $-$0.66 $\pm$ 0.29 & 4.8 & --- & $-$3.75 &  Y  \\ \\ 
246.37557 & 13.19678 & 19.49 $\pm$ 0.02 & 18.39 $\pm$ 0.01 & 20.89 $\pm$ 0.06 & $-$67.7 $\pm$ 5.8 & 0.54 $\pm$ 0.21 & $-$0.28 $\pm$ 0.18 & 10.0 & --- & $-$0.94 &  N  \\ \\ 
246.84517 & 13.6051 & 20.46 $\pm$ 0.03 & 19.84 $\pm$ 0.03 & 21.27 $\pm$ 0.09 & $-$17.0 $\pm$ 13.0 & $-$0.86 $\pm$ 0.47 & $-$1.76 $\pm$ 0.40 & 3.3 & --- & $-$1.12 &  N  \\ \\ 
246.75478 & 12.66005 & 20.12 $\pm$ 0.02 & 18.98 $\pm$ 0.02 & 21.82 $\pm$ 0.16 & 60.1 $\pm$ 4.3 & $-$1.58 $\pm$ 0.32 & $-$0.07 $\pm$ 0.32 & 13.5 & --- & $-$0.43 &  N  \\ \\ 
246.73328 & 12.5240 & 20.47 $\pm$ 0.03 & 19.43 $\pm$ 0.02 & 21.64 $\pm$ 0.16 & 32.8 $\pm$ 10.1 & $-$3.40 $\pm$ 0.39 & $-$1.66 $\pm$ 0.34 & 4.0 & --- & $-$1.57 &  N  \\ \\ 
247.94199 & 12.17424 & 19.91 $\pm$ 0.02 & 19.04 $\pm$ 0.02 & 21.10 $\pm$ 0.09 & $-$179.5 $\pm$ 13.7 & $-$1.80 $\pm$ 0.26 & $-$8.55 $\pm$ 0.22 & 5.5 & --- & $-$0.76 &  N  \\ \\ 
247.25868 & 12.37304 & 18.76 $\pm$ 0.01 & 17.58 $\pm$ 0.01 & 20.11 $\pm$ 0.05 & $-$41.9 $\pm$ 2.9 & $-$5.34 $\pm$ 0.19 & 1.98 $\pm$ 0.17 & 30.7 & --- & $-$1.50 &  N  \\ \\ 
247.3685 & 12.30693 & 20.59 $\pm$ 0.03 & 19.83 $\pm$ 0.03 & 21.87 $\pm$ 0.18 & 19.5 $\pm$ 11.3 & $-$2.23 $\pm$ 0.58 & $-$1.87 $\pm$ 0.51 & 3.6 & --- & $---$ &  N  \\ \\ 
247.09069 & 12.44679 & 20.14 $\pm$ 0.02 & 19.19 $\pm$ 0.02 & 21.38 $\pm$ 0.13 & $-$89.2 $\pm$ 5.1 & $-$2.02 $\pm$ 0.34 & $-$1.79 $\pm$ 0.35 & 8.7 & --- & $-$0.96 &  N  \\ \\ 
246.96076 & 12.66072 & 18.98 $\pm$ 0.01 & 17.63 $\pm$ 0.01 & 19.92 $\pm$ 0.04 & $-$119.4 $\pm$ 5.4 & $-$3.33 $\pm$ 0.16 & $-$6.12 $\pm$ 0.16 & 11.7 & --- & $---$ &  N  \\ \\ 
248.14307 & 11.9990 & 19.69 $\pm$ 0.02 & 18.56 $\pm$ 0.01 & 21.39 $\pm$ 0.12 & 8.3 $\pm$ 7.7 & 0.25 $\pm$ 0.19 & $-$4.60 $\pm$ 0.16 & 7.2 & --- & $-$0.38 &  N  \\ \\ 
246.9166 & 12.27304 & 20.63 $\pm$ 0.03 & 19.97 $\pm$ 0.03 & 21.56 $\pm$ 0.12 & 162.4 $\pm$ 8.4 & $-$9.39 $\pm$ 0.62 & 2.74 $\pm$ 0.62 & 4.2 & --- & $-$0.79 &  N  \\ \\ 
246.94378 & 12.82228 & 18.64 $\pm$ 0.01 & 17.38 $\pm$ 0.01 & 20.30 $\pm$ 0.05 & $-$4.6 $\pm$ 2.8 & $-$4.79 $\pm$ 0.12 & $-$0.57 $\pm$ 0.10 & 35.9 & --- & $-$0.92 &  N  \\ \\ 
246.89183 & 12.85064 & 20.11 $\pm$ 0.02 & 19.03 $\pm$ 0.02 & 21.67 $\pm$ 0.14 & $-$14.4 $\pm$ 6.9 & $-$10.06 $\pm$ 0.34 & $-$7.80 $\pm$ 0.34 & 7.1 & --- & $-$0.47 &  N  \\ \\ 
247.78112 & 12.04271 & 19.07 $\pm$ 0.01 & 18.18 $\pm$ 0.01 & 20.34 $\pm$ 0.05 & $-$43.4 $\pm$ 4.8 & $-$0.57 $\pm$ 0.15 & $-$0.63 $\pm$ 0.12 & 11.1 & --- & $-$0.53 &  N  \\ \\ 
247.54116 & 12.91777 & 19.61 $\pm$ 0.02 & 18.48 $\pm$ 0.01 & 20.40 $\pm$ 0.05 & 43.6 $\pm$ 3.5 & $-$0.40 $\pm$ 0.18 & $-$0.43 $\pm$ 0.15 & 12.8 & $-$3.0 $\pm$ 0.5 & $-$3.77 &  Y  \\ \\ 
246.6543 & 12.93348 & 20.01 $\pm$ 0.02 & 18.84 $\pm$ 0.02 & 21.79 $\pm$ 0.18 & $-$61.3 $\pm$ 12.2 & $-$2.84 $\pm$ 0.28 & 2.08 $\pm$ 0.28 & 5.7 & --- & $-$0.43 &  N  \\ \\ 
246.96346 & 12.70229 & 18.67 $\pm$ 0.01 & 17.39 $\pm$ 0.01 & 20.41 $\pm$ 0.06 & $-$9.7 $\pm$ 3.3 & 14.1 $\pm$ 0.11 & $-$9.86 $\pm$ 0.11 & 35.5 & --- & $-$0.80 &  N  \\ \\ 
246.87751 & 12.65129 & 20.15 $\pm$ 0.02 & 19.57 $\pm$ 0.02 & 20.93 $\pm$ 0.08 & $-$32.1 $\pm$ 10.2 & $-$3.91 $\pm$ 0.39 & $-$7.20 $\pm$ 0.37 & 10.2 & --- & $-$1.03 &  N  \\ \\ 
246.86191 & 12.59161 & 21.46 $\pm$ 0.08 & 20.48 $\pm$ 0.05 & 22.04 $\pm$ 0.17 & 6.5 $\pm$ 5.7 & $-$3.34 $\pm$ 1.08 & 0.03 $\pm$ 1.12 & 11.0 & --- & $---$ &  N  \\ \\ 
246.72922 & 12.55599 & 20.48 $\pm$ 0.03 & 19.74 $\pm$ 0.03 & 21.22 $\pm$ 0.11 & $-$27.7 $\pm$ 9.8 & $-$7.67 $\pm$ 0.51 & $-$15.18 $\pm$ 0.43 & 3.6 & --- & $-$2.10 &  N  \\ \\ 
246.76616 & 12.68591 & 20.14 $\pm$ 0.02 & 19.21 $\pm$ 0.02 & 21.15 $\pm$ 0.10 & $-$139.7 $\pm$ 11.7 & $-$2.68 $\pm$ 0.35 & $-$5.32 $\pm$ 0.34 & 8.4 & --- & $-$1.72 &  N  \\ \\ 
246.71304 & 12.56697 & 18.76 $\pm$ 0.01 & 17.45 $\pm$ 0.01 & 20.47 $\pm$ 0.06 & 47.5 $\pm$ 4.6 & $-$3.14 $\pm$ 0.49 & $-$3.32 $\pm$ 0.45 & 28.2 & --- & $-$0.89 &  N  \\ \\ 
246.64344 & 12.68448 & 19.69 $\pm$ 0.02 & 18.73 $\pm$ 0.02 & 21.06 $\pm$ 0.09 & 77.9 $\pm$ 9.3 & $-$9.38 $\pm$ 0.23 & $-$18.51 $\pm$ 0.21 & 7.0 & --- & $-$0.53 &  N  \\ \\ 
247.26691 & 12.30981 & 20.02 $\pm$ 0.02 & 18.98 $\pm$ 0.02 & 21.61 $\pm$ 0.16 & $-$1.9 $\pm$ 8.6 & $-$0.98 $\pm$ 0.28 & $-$8.82 $\pm$ 0.26 & 4.5 & --- & $-$0.20 &  N  \\ \\ 
248.58387 & 12.5407 & 20.20 $\pm$ 0.02 & 19.36 $\pm$ 0.02 & 21.32 $\pm$ 0.10 & 12.4 $\pm$ 7.2 & $-$1.63 $\pm$ 0.33 & 3.95 $\pm$ 0.26 & 4.3 & --- & $-$0.94 &  N  \\ \\ 
248.38652 & 11.94489 & 19.46 $\pm$ 0.02 & 18.51 $\pm$ 0.02 & 20.95 $\pm$ 0.07 & $-$70.1 $\pm$ 13.2 & 0.17 $\pm$ 0.16 & $-$7.85 $\pm$ 0.13 & 6.7 & --- & $-$0.12 &  N  \\ \\ 
247.18706 & 13.24547 & 19.87 $\pm$ 0.02 & 18.99 $\pm$ 0.02 & 20.60 $\pm$ 0.06 & 34.9 $\pm$ 9.9 & $-$4.23 $\pm$ 0.25 & $-$20.18 $\pm$ 0.22 & 10.4 & --- & $-$2.92 &  N  \\ \\ 
248.95752 & 11.67386 & 19.11 $\pm$ 0.01 & 18.23 $\pm$ 0.01 & 20.33 $\pm$ 0.04 & $-$92.6 $\pm$ 8.7 & $-$0.27 $\pm$ 0.13 & $-$0.04 $\pm$ 0.10 & 6.6 & --- & $-$0.79 &  N  \\ \\ 
246.86297 & 12.24227 & 19.59 $\pm$ 0.02 & 18.64 $\pm$ 0.02 & 20.83 $\pm$ 0.07 & 2.3 $\pm$ 7.7 & $-$0.40 $\pm$ 0.25 & $-$0.84 $\pm$ 0.22 & 5.5 & --- & $-$1.00 &  N  \\ \\ 
247.09492 & 13.14445 & 19.20 $\pm$ 0.01 & 17.93 $\pm$ 0.01 & 20.64 $\pm$ 0.07 & $-$115.9 $\pm$ 5.1 & $-$10.09 $\pm$ 0.12 & $-$1.63 $\pm$ 0.10 & 13.2 & --- & $-$1.49 &  N  \\ \\ 
247.05246 & 13.13164 & 19.69 $\pm$ 0.02 & 18.65 $\pm$ 0.01 & 21.23 $\pm$ 0.11 & $-$21.2 $\pm$ 14.6 & 0.38 $\pm$ 0.18 & $-$6.83 $\pm$ 0.16 & 4.9 & --- & $-$0.34 &  N  \\ \\ 
246.98739 & 13.1295 & 20.21 $\pm$ 0.02 & 19.13 $\pm$ 0.02 & 21.76 $\pm$ 0.18 & $-$75.7 $\pm$ 7.2 & $-$2.83 $\pm$ 0.27 & $-$2.65 $\pm$ 0.25 & 7.7 & --- & $-$0.48 &  N  \\ \\ 
247.01672 & 13.19916 & 20.57 $\pm$ 0.03 & 19.81 $\pm$ 0.03 & 21.22 $\pm$ 0.11 & $-$2.2 $\pm$ 9.0 & $-$2.99 $\pm$ 0.42 & $-$5.14 $\pm$ 0.35 & 4.2 & --- & $-$2.61 &  N  \\ \\ 
248.58885 & 11.90728 & 20.16 $\pm$ 0.02 & 19.38 $\pm$ 0.02 & 20.97 $\pm$ 0.09 & $-$116.9 $\pm$ 13.1 & $-$6.75 $\pm$ 0.32 & 1.23 $\pm$ 0.24 & 4.1 & --- & $-$1.95 &  N  \\ \\ 
246.73081 & 13.19528 & 20.27 $\pm$ 0.02 & 19.31 $\pm$ 0.02 & 21.25 $\pm$ 0.09 & $-$32.8 $\pm$ 9.2 & $-$0.48 $\pm$ 0.52 & $-$8.25 $\pm$ 0.47 & 5.3 & --- & $-$1.99 &  N  \\ \\ 
248.48765 & 11.8842 & 20.15 $\pm$ 0.03 & 19.16 $\pm$ 0.02 & 21.39 $\pm$ 0.12 & 21.3 $\pm$ 9.2 & 5.37 $\pm$ 0.28 & $-$7.26 $\pm$ 0.22 & 4.5 & --- & $-$1.14 &  N  \\ \\ 
246.68288 & 12.88727 & 18.99 $\pm$ 0.01 & 17.72 $\pm$ 0.01 & 20.69 $\pm$ 0.06 & $-$18.3 $\pm$ 2.8 & $-$13.58 $\pm$ 0.20 & $-$8.94 $\pm$ 0.20 & 21.5 & --- & $-$0.87 &  N  \\ \\ 
246.88408 & 12.59174 & 20.30 $\pm$ 0.03 & 19.38 $\pm$ 0.02 & 21.50 $\pm$ 0.12 & $-$30.6 $\pm$ 6.7 & $-$4.46 $\pm$ 0.40 & $-$5.21 $\pm$ 0.38 & 7.2 & --- & $-$1.04 &  N  \\ \\ 
246.88673 & 12.70166 & 20.40 $\pm$ 0.03 & 19.55 $\pm$ 0.02 & 21.86 $\pm$ 0.16 & $-$94.7 $\pm$ 9.4 & $-$9.56 $\pm$ 0.47 & $-$1.39 $\pm$ 0.42 & 4.1 & --- & $---$ &  N  \\ \\

\end{tabular}
\end{sideways}
\end{table*}

\newpage

\begin{table*}

\caption{Table 2 - Continued.
\label{table1}}

\setlength{\tabcolsep}{2.5pt}
\renewcommand{\arraystretch}{0.3}
\begin{sideways}
\begin{tabular}{cccccccccccccc}
\hline
RA (deg) & DEC (deg) & $g_0$ & $i_0$ & $CaHK_0$ & $\mathrm{v}_{r} (\kms)$ & $\mu_{\alpha}^{*}$ (mas.yr$^{-1}$) & $\mu_{\delta}$ (mas.yr$^{-1}$) &  S/N & [Fe/H]$^\mathrm{V}_\mathrm{1+2}$ & [Fe/H]$_\mathrm{Pristine}$ & Member\\

\hline

246.86038 & 12.65564 & 20.69 $\pm$ 0.03 & 19.81 $\pm$ 0.02 & 21.93 $\pm$ 0.16 & $-$98.8 $\pm$ 6.7 & $-$17.02 $\pm$ 0.52 & $-$3.90 $\pm$ 0.50 & 6.0 & --- & $-$0.68 &  N  \\ \\ 
246.82742 & 12.60486 & 19.89 $\pm$ 0.02 & 18.91 $\pm$ 0.02 & 21.04 $\pm$ 0.09 & 57.4 $\pm$ 7.5 & $-$2.10 $\pm$ 0.29 & $-$4.44 $\pm$ 0.30 & 6.9 & --- & $-$1.33 &  N  \\ \\ 
248.5845 & 12.59069 & 20.00 $\pm$ 0.02 & 19.26 $\pm$ 0.02 & 22.51 $\pm$ 0.24 & 74.7 $\pm$ 14.0 & $-$2.35 $\pm$ 0.28 & $-$6.53 $\pm$ 0.22 & 4.0 & --- & $---$ &  N  \\ \\ 
248.55005 & 12.5758 & 19.43 $\pm$ 0.02 & 18.27 $\pm$ 0.01 & 21.06 $\pm$ 0.08 & $-$86.1 $\pm$ 5.5 & 3.30 $\pm$ 0.17 & $-$6.11 $\pm$ 0.13 & 11.1 & --- & $-$0.66 &  N  \\ \\ 
247.09157 & 12.30991 & 19.13 $\pm$ 0.01 & 18.09 $\pm$ 0.01 & 20.49 $\pm$ 0.06 & $-$1.4 $\pm$ 3.1 & $-$4.88 $\pm$ 0.28 & 3.45 $\pm$ 0.29 & 20.9 & --- & $-$0.97 &  N  \\ \\ 
248.35298 & 12.51463 & 19.51 $\pm$ 0.02 & 18.18 $\pm$ 0.01 & 20.61 $\pm$ 0.06 & $-$37.8 $\pm$ 5.9 & $-$8.14 $\pm$ 0.12 & $-$0.23 $\pm$ 0.10 & 12.6 & --- & $-$3.45 &  N  \\ \\ 
247.03553 & 12.31056 & 20.77 $\pm$ 0.03 & 19.93 $\pm$ 0.03 & 21.54 $\pm$ 0.13 & $-$101.7 $\pm$ 7.8 & $-$4.51 $\pm$ 0.62 & $-$5.41 $\pm$ 0.63 & 3.5 & --- & $-$2.46 &  N  \\ \\ 
246.74741 & 12.44978 & 20.98 $\pm$ 0.04 & 20.29 $\pm$ 0.04 & 21.58 $\pm$ 0.15 & $-$37.6 $\pm$ 8.3 & 0.60 $\pm$ 0.66 & $-$1.06 $\pm$ 0.54 & 5.7 & --- & $-$2.53 &  ?  \\ \\ 
246.74694 & 12.31479 & 20.49 $\pm$ 0.03 & 19.46 $\pm$ 0.02 & 21.71 $\pm$ 0.16 & $-$116.6 $\pm$ 10.4 & $-$2.73 $\pm$ 0.42 & $-$13.1 $\pm$ 0.36 & 4.8 & --- & $-$1.40 &  N  \\ \\ 
246.69844 & 12.43831 & 19.18 $\pm$ 0.01 & 18.06 $\pm$ 0.01 & 20.66 $\pm$ 0.07 & 44.9 $\pm$ 3.5 & $-$1.07 $\pm$ 0.15 & $-$9.50 $\pm$ 0.13 & 14.3 & --- & $-$0.83 &  N  \\ \\ 
248.46686 & 12.0008 & 19.52 $\pm$ 0.02 & 18.55 $\pm$ 0.01 & 21.04 $\pm$ 0.08 & 27.7 $\pm$ 4.7 & 1.95 $\pm$ 0.18 & $-$0.05 $\pm$ 0.14 & 6.8 & --- & $-$0.13 &  N  \\ \\ 
247.19971 & 13.52296 & 19.47 $\pm$ 0.02 & 18.56 $\pm$ 0.01 & 20.79 $\pm$ 0.07 & $-$103.1 $\pm$ 8.7 & $-$5.88 $\pm$ 0.18 & $-$2.08 $\pm$ 0.14 & 7.7 & --- & $-$0.45 &  N  \\ \\ 
247.19206 & 13.44977 & 19.90 $\pm$ 0.02 & 18.75 $\pm$ 0.02 & 21.43 $\pm$ 0.11 & $-$55.4 $\pm$ 6.6 & $-$5.25 $\pm$ 0.20 & $-$12.62 $\pm$ 0.17 & 8.9 & --- & $-$0.78 &  N  \\ \\ 
247.24622 & 13.41695 & 19.93 $\pm$ 0.02 & 18.69 $\pm$ 0.02 & 21.69 $\pm$ 0.14 & $-$81.0 $\pm$ 6.5 & $-$0.28 $\pm$ 0.20 & $-$7.97 $\pm$ 0.17 & 8.1 & --- & $-$0.63 &  N  \\ \\ 
247.21778 & 13.57619 & 20.49 $\pm$ 0.03 & 19.66 $\pm$ 0.02 & 21.57 $\pm$ 0.14 & $-$141.8 $\pm$ 6.8 & $-$10.34 $\pm$ 0.40 & $-$12.98 $\pm$ 0.33 & 4.6 & --- & $-$1.02 &  N  \\ \\ 
247.88895 & 12.21281 & 19.37 $\pm$ 0.02 & 18.08 $\pm$ 0.01 & 21.01 $\pm$ 0.08 & 31.2 $\pm$ 5.1 & $-$5.15 $\pm$ 0.15 & $-$3.20 $\pm$ 0.12 & 14.0 & --- & $-$1.04 &  N  \\ \\ 
247.83506 & 12.03185 & 20.20 $\pm$ 0.02 & 19.35 $\pm$ 0.02 & 21.40 $\pm$ 0.11 & 12.7 $\pm$ 11.8 & $-$0.69 $\pm$ 0.35 & $-$1.91 $\pm$ 0.27 & 4.1 & --- & $-$0.66 &  N  \\ \\ 
247.4585 & 13.70662 & 19.67 $\pm$ 0.02 & 18.62 $\pm$ 0.02 & 20.54 $\pm$ 0.06 & 48.8 $\pm$ 6.1 & $-$0.20 $\pm$ 0.20 & $-$0.40 $\pm$ 0.17 & 8.7 & --- & $-$3.03 &  Y  \\ \\ 
248.58076 & 13.1713 & 19.03 $\pm$ 0.01 & 17.99 $\pm$ 0.01 & 20.57 $\pm$ 0.06 & $-$45.7 $\pm$ 9.9 & $-$9.11 $\pm$ 0.12 & $-$0.52 $\pm$ 0.10 & 7.3 & --- & $-$0.32 &  N  \\ \\ 
247.28533 & 13.26492 & 20.29 $\pm$ 0.03 & 19.11 $\pm$ 0.02 & 20.93 $\pm$ 0.08 & $-$132.1 $\pm$ 7.6 & $-$8.18 $\pm$ 0.21 & $-$13.54 $\pm$ 0.18 & 8.8 & --- & $---$ &  N  \\ \\ 
247.24962 & 13.33747 & 19.97 $\pm$ 0.02 & 19.08 $\pm$ 0.02 & 21.25 $\pm$ 0.10 & $-$100.0 $\pm$ 7.7 & $-$3.55 $\pm$ 0.24 & $-$3.44 $\pm$ 0.21 & 4.3 & --- & $-$0.52 &  N  \\ \\ 
246.85571 & 13.3048 & 20.50 $\pm$ 0.03 & 19.40 $\pm$ 0.02 & 21.93 $\pm$ 0.16 & 14.3 $\pm$ 6.1 & $-$0.82 $\pm$ 0.38 & $-$6.16 $\pm$ 0.35 & 5.3 & --- & $-$0.90 &  N  \\ \\ 
246.68136 & 13.33053 & 19.96 $\pm$ 0.02 & 18.78 $\pm$ 0.02 & 21.67 $\pm$ 0.15 & 0.2 $\pm$ 11.9 & $-$5.99 $\pm$ 0.26 & $-$0.21 $\pm$ 0.27 & 4.2 & --- & $-$0.54 &  N  \\ \\ 
248.57357 & 13.39772 & 18.58 $\pm$ 0.01 & 17.19 $\pm$ 0.01 & 20.41 $\pm$ 0.06 & $-$31.1 $\pm$ 4.2 & $-$2.59 $\pm$ 0.08 & $-$18.44 $\pm$ 0.07 & 13.2 & --- & $-$0.90 &  N  \\ \\ 
246.95402 & 12.63921 & 18.45 $\pm$ 0.01 & 17.12 $\pm$ 0.01 & 20.82 $\pm$ 0.08 & 3.7 $\pm$ 5.0 & $-$10.3 $\pm$ 0.09 & $-$9.15 $\pm$ 0.09 & 35.7 & --- & $---$ &  N  \\ \\ 
247.55111 & 13.20879 & 20.64 $\pm$ 0.03 & 19.97 $\pm$ 0.03 & 21.46 $\pm$ 0.11 & $-$90.1 $\pm$ 8.2 & $-$1.42 $\pm$ 0.42 & $-$1.42 $\pm$ 0.35 & 3.1 & --- & $-$1.40 &  N  \\ \\ 
246.87513 & 12.71872 & 18.76 $\pm$ 0.01 & 17.29 $\pm$ 0.01 & 20.57 $\pm$ 0.06 & $-$45.9 $\pm$ 4.0 & $-$6.27 $\pm$ 0.11 & 6.73 $\pm$ 0.11 & 28.3 & --- & $-$1.09 &  N  \\ \\ 
246.8359 & 12.71057 & 19.94 $\pm$ 0.02 & 18.70 $\pm$ 0.02 & 21.22 $\pm$ 0.10 & $-$45.9 $\pm$ 5.2 & $-$3.98 $\pm$ 0.24 & 0.93 $\pm$ 0.23 & 9.2 & --- & $-$1.95 &  N  \\ \\ 
246.79782 & 12.75613 & 19.71 $\pm$ 0.02 & 18.40 $\pm$ 0.01 & 21.24 $\pm$ 0.10 & $-$51.8 $\pm$ 4.5 & $-$1.81 $\pm$ 0.22 & $-$8.74 $\pm$ 0.21 & 11.4 & --- & $-$1.42 &  N  \\ \\ 
246.68838 & 12.63302 & 20.01 $\pm$ 0.03 & 19.11 $\pm$ 0.02 & 21.10 $\pm$ 0.09 & $-$185.5 $\pm$ 7.8 & $-$7.45 $\pm$ 0.33 & $-$19.2 $\pm$ 0.29 & 10.0 & --- & $-$1.34 &  N  \\ \\ 
247.25876 & 13.61103 & 20.14 $\pm$ 0.02 & 19.38 $\pm$ 0.02 & 21.23 $\pm$ 0.10 & $-$119.6 $\pm$ 7.3 & $-$5.68 $\pm$ 0.30 & $-$3.78 $\pm$ 0.25 & 5.4 & --- & $-$0.71 &  N  \\ \\ 
248.58281 & 12.69705 & 19.54 $\pm$ 0.02 & 18.31 $\pm$ 0.01 & 21.35 $\pm$ 0.12 & $-$10.0 $\pm$ 6.6 & 7.08 $\pm$ 0.15 & $-$9.65 $\pm$ 0.13 & 10.7 & --- & $-$0.57 &  N  \\ \\ 
246.08627 & 12.41528 & 19.78 $\pm$ 0.02 & 18.44 $\pm$ 0.01 & 21.36 $\pm$ 0.10 & $-$142.5 $\pm$ 5.4 & 0.24 $\pm$ 0.23 & 0.40 $\pm$ 0.22 & 10.4 & --- & $-$1.38 &  ?  \\ \\ 
246.89308 & 13.50018 & 20.32 $\pm$ 0.02 & 19.24 $\pm$ 0.02 & 21.57 $\pm$ 0.12 & 31.6 $\pm$ 9.0 & $-$6.40 $\pm$ 0.28 & $-$12.21 $\pm$ 0.25 & 5.4 & --- & $-$1.45 &  N  \\ \\ 
246.74212 & 13.48835 & 18.80 $\pm$ 0.01 & 17.40 $\pm$ 0.01 & 20.52 $\pm$ 0.07 & $-$19.1 $\pm$ 3.0 & $-$3.22 $\pm$ 0.09 & $-$2.08 $\pm$ 0.08 & 18.8 & --- & $-$1.15 &  N  \\ \\ 
247.47233 & 13.79749 & 20.47 $\pm$ 0.03 & 19.39 $\pm$ 0.02 & 21.84 $\pm$ 0.18 & $-$60.9 $\pm$ 13.0 & $-$4.74 $\pm$ 0.34 & 0.99 $\pm$ 0.30 & 3.4 & --- & $-$1.00 &  N  \\ \\ 
247.15456 & 13.78807 & 19.71 $\pm$ 0.02 & 18.97 $\pm$ 0.02 & 20.93 $\pm$ 0.09 & $-$127.6 $\pm$ 11.7 & $-$2.51 $\pm$ 0.25 & $-$2.06 $\pm$ 0.22 & 4.7 & --- & $-$0.23 &  N  \\ \\ 
247.12676 & 13.77042 & 19.90 $\pm$ 0.02 & 18.83 $\pm$ 0.02 & 21.53 $\pm$ 0.15 & 7.8 $\pm$ 6.1 & $-$1.34 $\pm$ 0.24 & $-$6.68 $\pm$ 0.19 & 8.2 & --- & $-$0.25 &  N  \\ \\ 
246.96895 & 13.72795 & 20.60 $\pm$ 0.03 & 19.48 $\pm$ 0.02 & 21.80 $\pm$ 0.16 & $-$154.3 $\pm$ 7.0 & $-$0.76 $\pm$ 0.37 & $-$23.09 $\pm$ 0.31 & 4.7 & --- & $-$1.92 &  N  \\ \\ 
246.91982 & 13.86001 & 19.80 $\pm$ 0.02 & 18.99 $\pm$ 0.02 & 20.78 $\pm$ 0.07 & $-$105.8 $\pm$ 7.3 & $-$8.66 $\pm$ 0.24 & 4.08 $\pm$ 0.22 & 5.5 & --- & $-$1.40 &  N  \\ \\ 
248.59121 & 12.96194 & 19.99 $\pm$ 0.02 & 18.78 $\pm$ 0.02 & 21.72 $\pm$ 0.16 & $-$4.2 $\pm$ 10.1 & 2.75 $\pm$ 0.21 & $-$11.55 $\pm$ 0.18 & 5.5 & --- & $-$0.59 &  N  \\ \\ 
246.77069 & 13.90807 & 19.27 $\pm$ 0.02 & 17.95 $\pm$ 0.01 & 20.66 $\pm$ 0.07 & 80.8 $\pm$ 4.2 & $-$3.15 $\pm$ 0.15 & $-$6.56 $\pm$ 0.13 & 20.0 & --- & $-$1.62 &  N  \\ \\ 
246.75267 & 13.85352 & 19.94 $\pm$ 0.02 & 18.72 $\pm$ 0.02 & 21.52 $\pm$ 0.14 & $-$28.6 $\pm$ 9.9 & $-$3.53 $\pm$ 0.22 & $-$4.82 $\pm$ 0.18 & 6.4 & --- & $-$0.87 &  N  \\ \\ 
246.62949 & 13.82289 & 19.36 $\pm$ 0.02 & 18.15 $\pm$ 0.01 & 21.03 $\pm$ 0.09 & $-$30.8 $\pm$ 3.6 & 0.58 $\pm$ 0.14 & $-$5.84 $\pm$ 0.12 & 15.9 & --- & $-$0.74 &  N  \\ \\ 
246.69948 & 13.80275 & 20.53 $\pm$ 0.03 & 19.84 $\pm$ 0.03 & 21.13 $\pm$ 0.10 & $-$86.0 $\pm$ 14.6 & $-$1.38 $\pm$ 0.48 & $-$1.08 $\pm$ 0.43 & 4.0 & --- & $-$2.52 &  N  \\ \\ 
246.66877 & 13.37787 & 20.08 $\pm$ 0.02 & 18.86 $\pm$ 0.02 & 21.49 $\pm$ 0.13 & $-$106.3 $\pm$ 9.0 & $-$3.64 $\pm$ 0.28 & $-$6.16 $\pm$ 0.29 & 7.0 & --- & $-$1.43 &  N  \\ \\ 
248.71753 & 12.63136 & 19.71 $\pm$ 0.02 & 18.67 $\pm$ 0.02 & 21.13 $\pm$ 0.12 & 6.1 $\pm$ 8.0 & $-$1.18 $\pm$ 0.19 & $-$10.98 $\pm$ 0.16 & 6.3 & --- & $-$0.81 &  N  \\ \\ 
248.7167 & 12.4921 & 20.11 $\pm$ 0.02 & 19.20 $\pm$ 0.02 & 21.48 $\pm$ 0.14 & 37.7 $\pm$ 7.0 & $-$4.54 $\pm$ 0.31 & $-$0.15 $\pm$ 0.24 & 6.8 & --- & $-$0.28 &  N  \\ \\ 
248.70384 & 12.49582 & 20.58 $\pm$ 0.03 & 19.75 $\pm$ 0.02 & 21.82 $\pm$ 0.18 & $-$59.2 $\pm$ 11.3 & 0.16 $\pm$ 0.44 & $-$11.94 $\pm$ 0.35 & 4.2 & --- & $-$0.48 &  N  \\ \\ 
246.66737 & 13.47796 & 19.06 $\pm$ 0.01 & 17.86 $\pm$ 0.01 & 20.57 $\pm$ 0.06 & 33.8 $\pm$ 3.2 & $-$0.50 $\pm$ 0.12 & $-$0.40 $\pm$ 0.11 & 14.8 & --- & $-$0.91 &  N  \\ \\ 
248.91509 & 12.19686 & 18.80 $\pm$ 0.01 & 17.46 $\pm$ 0.01 & 20.49 $\pm$ 0.05 & $-$13.5 $\pm$ 4.1 & $-$5.15 $\pm$ 0.09 & $-$2.03 $\pm$ 0.07 & 16.5 & --- & $-$1.09 &  N  \\ \\ 
248.84188 & 12.35783 & 19.05 $\pm$ 0.01 & 17.69 $\pm$ 0.01 & 20.79 $\pm$ 0.07 & $-$103.4 $\pm$ 4.1 & $-$5.67 $\pm$ 0.64 & $-$2.20 $\pm$ 0.53 & 28.4 & --- & $-$1.03 &  N  \\ \\ 
248.91006 & 12.27981 & 20.33 $\pm$ 0.03 & 19.36 $\pm$ 0.02 & 21.65 $\pm$ 0.13 & $-$173.7 $\pm$ 9.9 & $-$5.92 $\pm$ 0.33 & $-$10.56 $\pm$ 0.28 & 7.7 & --- & $-$0.81 &  N  \\ \\ 
248.84028 & 12.30854 & 20.75 $\pm$ 0.03 & 19.80 $\pm$ 0.03 & 21.86 $\pm$ 0.17 & $-$2.2 $\pm$ 6.9 & $-$0.13 $\pm$ 0.43 & $-$18.92 $\pm$ 0.34 & 3.6 & --- & $-$1.43 &  N  \\ \\ 
248.72804 & 12.25439 & 20.10 $\pm$ 0.03 & 19.08 $\pm$ 0.02 & 21.44 $\pm$ 0.10 & 6.9 $\pm$ 7.9 & $-$4.84 $\pm$ 0.26 & $-$6.29 $\pm$ 0.20 & 8.1 & --- & $-$0.97 &  N  \\ \\

\end{tabular}
\end{sideways}
\end{table*}

\newpage

\begin{table*}
\caption{Table 2 - Continued.}
\label{table1}

\setlength{\tabcolsep}{2.5pt}
\renewcommand{\arraystretch}{0.3}
\begin{sideways}
\begin{tabular}{cccccccccccccc}
\hline
RA (deg) & DEC (deg) & $g_0$ & $i_0$ & $CaHK_0$ & $\mathrm{v}_{r} (\kms)$ & $\mu_{\alpha}^{*}$ (mas.yr$^{-1}$) & $\mu_{\delta}$ (mas.yr$^{-1}$) & S/N & [Fe/H]$^\mathrm{V}_\mathrm{1+2}$ & [Fe/H]$_\mathrm{Pristine}$ & Member\\

\hline

248.52523 & 12.32225 & 19.78 $\pm$ 0.02 & 18.89 $\pm$ 0.02 & 21.16 $\pm$ 0.09 & $-$17.2 $\pm$ 8.6 & 0.01 $\pm$ 0.23 & $-$7.70 $\pm$ 0.18 & 4.7 & --- & $-$0.24 &  N  \\ \\ 
247.0681 & 12.58077 & 20.05 $\pm$ 0.02 & 19.10 $\pm$ 0.02 & 21.52 $\pm$ 0.15 & $-$6.5 $\pm$ 5.8 & $-$0.38 $\pm$ 0.36 & $-$0.51 $\pm$ 0.35 & 8.9 & --- & $-$0.21 &  N  \\ \\ 
246.9519 & 12.54664 & 20.62 $\pm$ 0.03 & 19.94 $\pm$ 0.03 & 21.69 $\pm$ 0.16 & 21.6 $\pm$ 11.3 & 1.19 $\pm$ 0.68 & $-$16.94 $\pm$ 0.61 & 5.7 & --- & $-$0.23 &  N  \\ \\ 
248.73247 & 12.63949 & 19.76 $\pm$ 0.02 & 18.80 $\pm$ 0.02 & 21.40 $\pm$ 0.15 & $-$58.8 $\pm$ 7.6 & 0.14 $\pm$ 0.20 & $-$0.54 $\pm$ 0.16 & 5.9 & --- & $---$ &  N  \\ \\ 
246.50552 & 13.50975 & 19.81 $\pm$ 0.02 & 19.03 $\pm$ 0.02 & 20.78 $\pm$ 0.05 & 6.3 $\pm$ 8.7 & $-$1.08 $\pm$ 0.29 & $-$1.04 $\pm$ 0.31 & 6.7 & --- & $-$1.28 &  N  \\ \\ 
248.84095 & 11.93982 & 19.71 $\pm$ 0.02 & 18.65 $\pm$ 0.01 & 21.00 $\pm$ 0.08 & 0.7 $\pm$ 6.4 & $-$0.60 $\pm$ 0.19 & $-$1.84 $\pm$ 0.15 & 9.2 & --- & $-$1.23 &  N  \\ \\ 
248.73755 & 12.10452 & 18.86 $\pm$ 0.01 & 17.88 $\pm$ 0.01 & 20.27 $\pm$ 0.05 & 9.1 $\pm$ 3.4 & $-$2.00 $\pm$ 0.10 & $-$14.28 $\pm$ 0.08 & 18.2 & --- & $-$0.52 &  N  \\ \\ 
248.7234 & 12.07086 & 20.90 $\pm$ 0.05 & 20.42 $\pm$ 0.05 & 22.22 $\pm$ 0.24 & $-$154.9 $\pm$ 11.4 & $-$4.37 $\pm$ 0.52 & $-$5.11 $\pm$ 0.45 & 3.2 & --- & $---$ &  N  \\ \\ 
248.75347 & 12.09482 & 20.20 $\pm$ 0.02 & 19.15 $\pm$ 0.02 & 21.45 $\pm$ 0.11 & $-$27.2 $\pm$ 9.2 & $-$1.99 $\pm$ 0.28 & $-$7.75 $\pm$ 0.22 & 9.3 & --- & $-$1.37 &  N  \\ \\ 
247.08649 & 12.6308 & 20.09 $\pm$ 0.02 & 19.08 $\pm$ 0.02 & 21.39 $\pm$ 0.12 & 45.5 $\pm$ 7.0 & 1.66 $\pm$ 0.33 & $-$4.65 $\pm$ 0.31 & 7.1 & --- & $-$1.08 &  N  \\ \\ 
248.78168 & 11.71213 & 20.03 $\pm$ 0.02 & 18.83 $\pm$ 0.02 & 21.43 $\pm$ 0.11 & $-$18.9 $\pm$ 12.0 & 3.46 $\pm$ 0.19 & $-$7.15 $\pm$ 0.16 & 5.5 & --- & $-$1.40 &  N  \\ \\ 
247.01658 & 12.58344 & 19.91 $\pm$ 0.02 & 18.94 $\pm$ 0.02 & 21.26 $\pm$ 0.11 & 14.3 $\pm$ 5.9 & $-$4.39 $\pm$ 0.29 & $-$4.00 $\pm$ 0.28 & 5.4 & --- & $-$0.71 &  N  \\ \\ 
246.99728 & 12.74368 & 20.24 $\pm$ 0.02 & 19.43 $\pm$ 0.02 & 21.61 $\pm$ 0.15 & $-$110.9 $\pm$ 4.1 & $-$2.32 $\pm$ 0.44 & $-$9.32 $\pm$ 0.38 & 7.0 & --- & $---$ &  N  \\ \\ 
247.03631 & 12.0440 & 19.87 $\pm$ 0.02 & 18.66 $\pm$ 0.02 & 21.44 $\pm$ 0.12 & $-$46.5 $\pm$ 5.5 & $-$4.04 $\pm$ 0.25 & $-$1.22 $\pm$ 0.24 & 10.8 & --- & $-$0.88 &  N  \\ \\ 
248.84067 & 11.87536 & 19.72 $\pm$ 0.02 & 18.87 $\pm$ 0.02 & 20.67 $\pm$ 0.06 & $-$145.0 $\pm$ 12.3 & $-$17.14 $\pm$ 0.20 & $-$15.86 $\pm$ 0.16 & 5.9 & --- & $-$1.63 &  N  \\ \\ 
248.9065 & 12.9504 & 20.04 $\pm$ 0.02 & 19.01 $\pm$ 0.02 & 21.66 $\pm$ 0.13 & 39.4 $\pm$ 9.4 & $-$0.69 $\pm$ 0.26 & $-$1.39 $\pm$ 0.23 & 4.1 & --- & $-$0.12 &  N  \\ \\ 
247.13655 & 12.73514 & 19.51 $\pm$ 0.02 & 18.62 $\pm$ 0.01 & 20.89 $\pm$ 0.09 & $-$0.3 $\pm$ 5.7 & $-$6.45 $\pm$ 0.22 & $-$0.28 $\pm$ 0.21 & 11.3 & --- & $-$0.23 &  N  \\ \\ 
248.90786 & 12.54507 & 20.05 $\pm$ 0.03 & 19.22 $\pm$ 0.02 & 20.83 $\pm$ 0.07 & 72.8 $\pm$ 11.2 & 0.84 $\pm$ 0.29 & $-$4.89 $\pm$ 0.24 & 5.0 & --- & $-$2.30 &  N  \\ \\ 
248.85394 & 12.71637 & 20.75 $\pm$ 0.03 & 19.79 $\pm$ 0.03 & 21.46 $\pm$ 0.11 & $-$24.6 $\pm$ 7.1 & $-$9.28 $\pm$ 0.51 & 4.14 $\pm$ 0.45 & 4.9 & --- & $-$3.70 &  N  \\ \\ 
248.77785 & 12.59378 & 20.53 $\pm$ 0.03 & 19.75 $\pm$ 0.03 & 21.24 $\pm$ 0.10 & $-$14.7 $\pm$ 11.7 & $-$1.89 $\pm$ 0.39 & $-$1.16 $\pm$ 0.32 & 3.1 & --- & $-$2.48 &  N  \\ \\ 
248.77866 & 12.69494 & 20.10 $\pm$ 0.02 & 19.18 $\pm$ 0.02 & 21.51 $\pm$ 0.16 & $-$19.5 $\pm$ 14.0 & $-$0.81 $\pm$ 0.27 & $-$10.37 $\pm$ 0.22 & 6.6 & --- & $-$0.22 &  N  \\ \\ 
248.76673 & 12.70066 & 20.23 $\pm$ 0.02 & 19.15 $\pm$ 0.02 & 21.48 $\pm$ 0.11 & $-$80.4 $\pm$ 7.1 & $-$6.90 $\pm$ 0.27 & $-$12.16 $\pm$ 0.22 & 7.7 & --- & $-$1.47 &  N  \\ \\ 
248.67474 & 12.59263 & 20.35 $\pm$ 0.03 & 19.26 $\pm$ 0.02 & 21.89 $\pm$ 0.15 & $-$13.3 $\pm$ 6.3 & $-$1.96 $\pm$ 0.30 & $-$10.58 $\pm$ 0.24 & 3.5 & --- & $-$0.64 &  N  \\ \\ 
246.34015 & 12.9162 & 19.91 $\pm$ 0.02 & 19.11 $\pm$ 0.02 & 20.95 $\pm$ 0.07 & $-$22.0 $\pm$ 6.7 & 0.15 $\pm$ 0.29 & $-$0.37 $\pm$ 0.26 & 5.2 & --- & $-$1.05 &  N  \\ \\ 
248.86753 & 12.37052 & 20.12 $\pm$ 0.02 & 19.13 $\pm$ 0.02 & 21.40 $\pm$ 0.11 & $-$96.8 $\pm$ 5.5 & 0.19 $\pm$ 0.26 & $-$1.53 $\pm$ 0.21 & 6.5 & --- & $-$1.04 &  N  \\ \\ 
248.8243 & 12.33862 & 20.03 $\pm$ 0.02 & 18.84 $\pm$ 0.02 & 21.60 $\pm$ 0.13 & $-$25.8 $\pm$ 5.7 & $-$1.17 $\pm$ 0.22 & $-$4.78 $\pm$ 0.18 & 7.3 & --- & $-$0.80 &  N  \\ \\ 
248.50171 & 12.41356 & 19.29 $\pm$ 0.02 & 18.35 $\pm$ 0.01 & 20.60 $\pm$ 0.06 & 7.1 $\pm$ 12.5 & $-$1.74 $\pm$ 0.33 & $-$7.39 $\pm$ 0.27 & 6.7 & --- & $-$0.65 &  N  \\ \\ 
248.88532 & 12.01262 & 19.14 $\pm$ 0.02 & 17.68 $\pm$ 0.01 & 20.88 $\pm$ 0.07 & 4.6 $\pm$ 3.0 & $-$3.92 $\pm$ 0.11 & $-$7.39 $\pm$ 0.08 & 18.7 & --- & $-$1.20 &  N  \\ \\ 
248.8507 & 11.99434 & 19.57 $\pm$ 0.02 & 18.58 $\pm$ 0.02 & 20.94 $\pm$ 0.07 & $-$118.8 $\pm$ 8.7 & $-$11.57 $\pm$ 0.16 & $-$7.94 $\pm$ 0.13 & 9.6 & --- & $-$0.69 &  N  \\ \\ 
248.90659 & 12.19325 & 19.74 $\pm$ 0.02 & 18.58 $\pm$ 0.02 & 21.07 $\pm$ 0.08 & 13.8 $\pm$ 5.4 & 7.14 $\pm$ 0.16 & $-$17.43 $\pm$ 0.14 & 10.4 & --- & $-$1.53 &  N  \\ \\ 
248.62262 & 12.20119 & 19.91 $\pm$ 0.02 & 19.06 $\pm$ 0.02 & 21.11 $\pm$ 0.08 & $-$4.3 $\pm$ 8.6 & $-$1.32 $\pm$ 0.23 & $-$11.64 $\pm$ 0.18 & 3.9 & --- & $-$0.65 &  N  \\ \\ 
248.66906 & 12.1607 & 20.26 $\pm$ 0.02 & 19.33 $\pm$ 0.02 & 21.51 $\pm$ 0.12 & $-$96.2 $\pm$ 13.4 & $-$8.47 $\pm$ 0.28 & $-$6.32 $\pm$ 0.23 & 5.3 & --- & $-$0.83 &  N  \\ \\ 
246.94712 & 12.75213 & 20.05 $\pm$ 0.02 & 18.88 $\pm$ 0.02 & 21.42 $\pm$ 0.13 & 14.7 $\pm$ 3.4 & $-$8.32 $\pm$ 0.29 & $-$2.07 $\pm$ 0.29 & 13.5 & --- & $-$1.38 &  N  \\ \\ 
248.84995 & 13.20057 & 18.59 $\pm$ 0.01 & 17.30 $\pm$ 0.01 & 19.85 $\pm$ 0.04 & $-$4.0 $\pm$ 10.7 & $-$13.26 $\pm$ 0.08 & 8.80 $\pm$ 0.07 & 11.2 & --- & $-$2.14 &  N  \\ \\ 
248.90558 & 13.05926 & 19.63 $\pm$ 0.02 & 17.49 $\pm$ 0.01 & 20.74 $\pm$ 0.09 & 42.7 $\pm$ 4.6 & $-$6.81 $\pm$ 0.10 & $-$7.38 $\pm$ 0.08 & 20.6 & --- & $---$ &  N  \\ \\ 
247.03244 & 13.73306 & 20.43 $\pm$ 0.03 & 19.57 $\pm$ 0.03 & 21.25 $\pm$ 0.12 & $-$113.3 $\pm$ 9.3 & $-$1.78 $\pm$ 0.36 & $-$26.46 $\pm$ 0.30 & 4.5 & --- & $-$2.32 &  N  \\ \\ 
248.55112 & 13.0871 & 19.25 $\pm$ 0.02 & 18.03 $\pm$ 0.01 & 20.97 $\pm$ 0.10 & $-$150.2 $\pm$ 10.7 & 2.77 $\pm$ 0.13 & $-$7.28 $\pm$ 0.11 & 10.6 & --- & $-$0.66 &  N  \\ \\ 
247.2214 & 13.65181 & 20.22 $\pm$ 0.03 & 19.32 $\pm$ 0.02 & 21.33 $\pm$ 0.11 & $-$172.9 $\pm$ 10.7 & $-$5.87 $\pm$ 0.29 & $-$12.11 $\pm$ 0.24 & 7.1 & --- & $-$1.26 &  N  \\ \\ 
247.1671 & 13.54161 & 19.88 $\pm$ 0.02 & 18.60 $\pm$ 0.01 & 21.44 $\pm$ 0.12 & $-$79.6 $\pm$ 4.4 & $-$6.13 $\pm$ 0.19 & 7.27 $\pm$ 0.16 & 9.7 & --- & $-$1.22 &  N  \\ \\ 
247.08076 & 13.56689 & 19.85 $\pm$ 0.02 & 18.92 $\pm$ 0.02 & 21.05 $\pm$ 0.11 & $-$175.3 $\pm$ 8.1 & $-$5.30 $\pm$ 0.23 & $-$5.02 $\pm$ 0.19 & 8.8 & --- & $-$0.99 &  N  \\ \\ 
248.9193 & 12.76459 & 19.79 $\pm$ 0.02 & 18.84 $\pm$ 0.02 & 20.97 $\pm$ 0.08 & $-$19.4 $\pm$ 8.2 & $-$1.43 $\pm$ 0.23 & $-$3.29 $\pm$ 0.18 & 7.5 & --- & $-$1.20 &  N  \\ \\ 
248.93277 & 12.00676 & 20.11 $\pm$ 0.02 & 19.40 $\pm$ 0.02 & 20.97 $\pm$ 0.06 & $-$244.9 $\pm$ 12.2 & $-$0.90 $\pm$ 0.30 & $-$0.85 $\pm$ 0.22 & 3.6 & --- & $-$1.40 &  N  \\ \\ 
247.17602 & 13.02714 & 19.72 $\pm$ 0.02 & 18.61 $\pm$ 0.01 & 21.05 $\pm$ 0.09 & $-$3.8 $\pm$ 6.5 & $-$2.49 $\pm$ 0.18 & $-$5.43 $\pm$ 0.16 & 12.5 & --- & $-$1.27 &  N  \\ \\ 
247.6062 & 12.33472 & 19.51 $\pm$ 0.02 & 18.45 $\pm$ 0.02 & 20.27 $\pm$ 0.05 & 235.1 $\pm$ 13.0 & 0.43 $\pm$ 0.61 & $-$0.49 $\pm$ 0.51 & 9.0 & --- & $-$3.71 &  N  \\ \\ 
247.03182 & 13.01717 & 18.47 $\pm$ 0.01 & 17.14 $\pm$ 0.01 & 20.09 $\pm$ 0.05 & $-$26.2 $\pm$ 11.1 & $-$3.20 $\pm$ 0.15 & $-$1.52 $\pm$ 0.13 & 38.8 & --- & $-$1.21 &  N  \\ \\ 
247.52398 & 12.2559 & 20.70 $\pm$ 0.03 & 19.95 $\pm$ 0.03 & 21.23 $\pm$ 0.10 & $-$57.0 $\pm$ 5.1 & $-$4.88 $\pm$ 0.57 & $-$9.70 $\pm$ 0.48 & 6.5 & --- & $-$3.37 &  N  \\ \\ 
248.86556 & 12.85928 & 19.77 $\pm$ 0.02 & 18.46 $\pm$ 0.01 & 21.32 $\pm$ 0.10 & $-$55.4 $\pm$ 7.8 & $-$2.74 $\pm$ 0.15 & $-$4.24 $\pm$ 0.13 & 11.2 & --- & $-$1.35 &  N  \\ \\ 
248.85031 & 12.50588 & 20.07 $\pm$ 0.03 & 19.16 $\pm$ 0.02 & 21.23 $\pm$ 0.10 & $-$0.6 $\pm$ 6.9 & $-$6.46 $\pm$ 0.29 & $-$4.05 $\pm$ 0.23 & 6.1 & --- & $-$1.09 &  N  \\ \\ 
248.80693 & 12.50851 & 20.38 $\pm$ 0.03 & 19.42 $\pm$ 0.03 & 21.66 $\pm$ 0.16 & $-$185.5 $\pm$ 7.6 & $-$3.00 $\pm$ 0.46 & $-$8.04 $\pm$ 0.39 & 5.8 & --- & $-$0.92 &  N  \\ \\ 
248.63792 & 12.49248 & 19.98 $\pm$ 0.02 & 18.79 $\pm$ 0.02 & 21.53 $\pm$ 0.13 & 28.5 $\pm$ 11.1 & $-$8.27 $\pm$ 0.22 & $-$7.02 $\pm$ 0.18 & 6.0 & --- & $-$0.84 &  N  \\ \\ 
248.37505 & 12.37005 & 20.03 $\pm$ 0.02 & 18.82 $\pm$ 0.02 & 21.54 $\pm$ 0.12 & 15.2 $\pm$ 5.9 & $-$1.65 $\pm$ 0.22 & $-$2.46 $\pm$ 0.17 & 4.2 & --- & $-$1.01 &  N  \\ \\ 
248.39447 & 13.15989 & 20.02 $\pm$ 0.03 & 18.85 $\pm$ 0.02 & 21.47 $\pm$ 0.12 & 20.3 $\pm$ 12.7 & 7.06 $\pm$ 0.20 & $-$14.83 $\pm$ 0.16 & 4.0 & --- & $-$1.04 &  N  \\ \\ 

\end{tabular}
\end{sideways}
\end{table*}

\begin{table*}

\caption{Probable misidentified member stars from the literature. The mention ``CaHK'' in the last column indicates that the CaHK is decisive in the decision-making, ``PM'' that the proper motion is.
\label{table2}}

\setlength{\tabcolsep}{2.5pt}
\renewcommand{\arraystretch}{0.3}
\begin{tabular}{cccccccccccccc}
\hline
RA (deg) & DEC (deg) & $g_0$ & $i_0$ & $CaHK_0$ & $\mathrm{v}_{r} (\kms)$ & $\mu_{\alpha}^{*}$ (mas.yr$^{-1}$) & $\mu_{\delta}$ (mas.yr$^{-1}$) & [Fe/H]$_\mathrm{spectro}$  \\

\hline

247.79628 & 12.78814 & 19.39 $\pm$ 0.02 & 18.15 $\pm$ 0.01 & 21.17 $\pm$ 0.10 & 30.7 $\pm$ 2.2 & $-$17.57 $\pm$ 0.14 & $-$17.67 $\pm$ 0.12 & $-$1.5 $\pm$ 0.2 & PM \\ \\ 
247.9701 & 12.7466 & 21.12 $\pm$ 0.04 & 21.56 $\pm$ 0.10 & 21.29 $\pm$ 0.13 & 7.0 $\pm$ 11.0 & --- & --- & --- & CaHK \\ \\ 
247.7754 & 12.7951 & 21.09 $\pm$ 0.04 & 21.53 $\pm$ 0.12 & 21.56 $\pm$ 0.17 & 61.0 $\pm$ 13.0 & --- & --- & --- & CaHK \\ \\ 
247.7699 & 12.7390 & 21.12 $\pm$ 0.04 & 21.50 $\pm$ 0.11 & 21.23 $\pm$ 0.12 & 40.0 $\pm$ 14.0 & --- & --- & --- & CaHK \\ \\ 
247.7605 & 12.8298 & 21.16 $\pm$ 0.04 & 21.53 $\pm$ 0.09 & 21.40 $\pm$ 0.13 & 31.0 $\pm$ 12.0 & --- & --- & --- & CaHK \\ \\ 
247.7493 & 12.7674 & 20.21 $\pm$ 0.02 & 20.26 $\pm$ 0.03 & 20.10 $\pm$ 0.04 & 64.0 $\pm$ 8.0 & 0.01 $\pm$ 0.62 & $-$0.60 $\pm$ 0.51 & --- & CaHK \\ \\ 
247.7242 & 12.8450 & 20.64 $\pm$ 0.03 & 20.25 $\pm$ 0.03 & 20.91 $\pm$ 0.10 & 65.0 $\pm$ 10.0 & $-$0.50 $\pm$ 0.68 & $-$0.11 $\pm$ 0.56 & --- & CaHK \\ \\ 
247.5792 & 12.8839 & 21.08 $\pm$ 0.04 & 21.04 $\pm$ 0.08 & 21.40 $\pm$ 0.11 & 80.0 $\pm$ 12.0 & --- & --- & --- & CaHK \\ \\ 
247.7566 & 12.7861 & 21.94 $\pm$ 0.08 & 21.37 $\pm$ 0.08 & 22.50 $\pm$ 0.05 & 41.4 $\pm$ 3.1 & --- & --- & --- & CaHK \\ \\ 
247.7758 & 12.8086 & 22.25 $\pm$ 0.10 & 21.40 $\pm$ 0.10 & 23.42 $\pm$ 0.10 & 48.5 $\pm$ 3.5 & --- & --- & --- & CaHK \\ \\ 
247.6033 & 12.8864 & 20.83 $\pm$ 0.03 & 19.89 $\pm$ 0.03 & 22.03 $\pm$ 0.19 & 53.1 $\pm$ 2.5 & 3.82 $\pm$ 0.61 & $-$20.14 $\pm$ 0.50 & $-$1.3 $\pm$ 0.1 & PM \\ \\ 
247.7911 & 12.7951 & 20.80 $\pm$ 0.03 & 20.04 $\pm$ 0.03 & 21.55 $\pm$ 0.15 & 41.0 $\pm$ 6.2 & $-$0.37 $\pm$ 0.50 & $-$0.44 $\pm$ 0.44 & --- & CaHK \\ \\

\end{tabular}
\end{table*}

\newpage

\newcommand{\mnras}{MNRAS}
\newcommand{\pasa}{PASA}
\newcommand{\nat}{Nature}
\newcommand{\araa}{ARAA}
\newcommand{\aj}{AJ}
\newcommand{\apj}{ApJ}
\newcommand{\apjl}{ApJ}
\newcommand{\apjs}{ApJSupp}
\newcommand{\aap}{A\&A}
\newcommand{\aaps}{A\&ASupp}
\newcommand{\pasp}{PASP}
\newcommand{\pasj}{PASJ}


\clearpage

\end{document}